\documentclass[aps,pre,  showpacs,superscriptaddress,amsmath,amssymb,color]{revtex4}
\usepackage{graphicx,subfigure}
\bibstyle{apsrev4-1}
\usepackage{color}

\usepackage{mathrsfs}
\providecommand\bcdot{\boldsymbol{\cdot}}

\newsavebox{\astrutbox}
\sbox{\astrutbox}{\rule[-5pt]{0pt}{20pt}}

\begin{document}
\title{Two-fluid kinetic theory for   dilute polymer solutions}
\author{Shiwani Singh}
\email[]{shiwani.singh@warwick.ac.uk}
 \affiliation{Mathematics Institute, University of Warwick, Coventry CV4 7AL, United Kingdom}
\affiliation{Engineering Mechanics Unit, JNCASR, Jakkur, Bangalore 560064, India}

\author{Ganesh Subramanian}
\email[]{sganesh@jncasr.ac.in}
\affiliation{Engineering Mechanics Unit, JNCASR, Jakkur, Bangalore
 560064, India}
 \author{Santosh Ansumali}
 \email[]{ansumali@jncasr.ac.in}
 \affiliation{Engineering Mechanics Unit, JNCASR, Jakkur, Bangalore
 560064, India}

 \begin{abstract}
We provide a Boltzmann-type kinetic description for   dilute polymer solutions based on two-fluid theory.  
This Boltzmann-type description  uses a quasi-equilibrium based relaxation
mechanism to model collisions   between a polymer dumbbell
and a solvent molecule. The model reproduces the desired macroscopic equations for
the polymer-solvent mixture. The proposed kinetic scheme leads to a numerical algorithm which is along the lines of the lattice Boltzmann method. Finally, the algorithm is applied to describe the evolution of a perturbed Kolmogorov flow profile, whereby  we  recover the major elastic effect exhibited by a polymer solution, specifically, the suppression of the original inertial instability.
 \end{abstract}
\maketitle
\section{Introduction}

The  numerical modeling of flows of polymeric liquids is often done via micro-macro simulations where one couples a continuum Navier-Stokes solver with a microscopic solver for the polymer dynamics. One of the simplest micro-mechanical approaches for modeling dilute polymer solutions in this manner is to treat them as a suspension of non-interacting elastic dumbbells immersed in a Newtonian solvent
\cite{bird_vol_1,bird_vol_2,larson1988constitutive}. For Hookean dumbbells, it is also possible to obtain a macroscopic constitutive equation for the stress tensor in closed form\,(the Oldroyd-B model\cite{bird_vol_2,larson1988constitutive}), and thereby, have a purely continuum model for flow behavior. The distinct advantage of using a microscopic approach for the polymer is that it is possible to solve for the flow even in circumstances which preclude the derivation of a closed-form constitutive equation in terms of macroscopic variables. The latter is the case for a suspension of FENE (finitely extensible nonlinearly elastic) dumbbells\cite{keunings1997peterlin,lielens1998new,lielens1999fene}. In most of the micro-macro approaches, the macroscopic flow solver, which solves the equations of motion using standard numerical techniques\,(finite difference or finite element), is coupled with microscopic Brownian dynamics (BD) simulations where one solves a large system of Langevin equations for the actual polymer molecules\,(the so-called CONFESSIT approach), or equivalent Brownian configuration fields, to obtain ensemble-averaged configuration statistics\cite{Lasoottinger1993,feiglottinger1995,ottinger1996stochastic,hulsen1997simulation}. Thus, in this approach, while the kinetic theory of polymer dynamics, based on an underlying Fokker-Planck equation, is considered, the solvent is still treated at the continuum level. In recent years, kinetic-theory-based solvers such as the lattice-Boltzmann (LB) formulation have emerged as an alternative to direct solvers of Navier-Stokes equations \cite{chen_annual_rev,succi_book,aidun2010lattice}. Due to the efficiency of such solvers, instead of macro-micro coupling, meso-micro coupling, wherein mesoscopic solvent flow solvers\,(LB, DPD, MPCD) replace the macroscopic flow solvers, is increasingly being advocated\cite{ahlrichs1998lattice,jendrejack2004shear,pham2009implicit,ahlrichs1999simulation,jain2012optimization}. In many of these cases, the polymer-solvent coupling is achieved by a simple dissipative ansatz\,\cite{ahlrichs1998lattice,ahlrichs1999simulation}. It would be natural to provide a kinetic theory framework where, along the lines of the original and classical case of gaseous mixtures\cite{sirovich1,sirovich2}, the solvent and solute are both modeled at the mesoscopic level. In the present case, this would imply a Boltzmann\,(or BGK)-based description of the solvent and a Fokker-Planck type description of the polymer, and the aforementioned dissipative coupling would then emerge naturally in the resulting moment equations. A number of discrete algorithms exist where some version of polymer kinetic or constitutive equation is solved along with an LB solver for the fluid\cite{Onishi2005,Malapinas2010,Gupta2015}. To the best of our knowledge, however, a Boltzmann (or Fokker-Planck) type kinetic equation, which can describe the two-fluid dynamics of a polymer-solvent mixture, does not exist.

Moreover, any attempt to extend the original Boltzmann mixture theory has to consider fundamental issues absent in the kinetic theories for mixtures of structureless particles\cite{sirovich1,sirovich2}. For example, modelling the polymer-solvent mixture needs one to account for the internal microstructure of the polymer molecules. It is the existence of these internal configurational degrees of freedom that lead to the characteristic entropic elasticity associated with flexible polymer chains. The momentum balance for a polymer solution may be written in the form\citep{milner1991hydrodynamics}:
\begin{align}
 \begin{split}
  \frac{\partial {\pmb J}}{\partial t}
+\frac{\partial }{\partial {\pmb r}}(\rho\,{\pmb U}{\pmb U})&=-\frac{\partial }{\partial {\pmb r}}(p+P_{\rm P}^{\rm osmotic})+ \frac{\partial }{\partial {\pmb r}}\cdot\left({\pmb \Pi}_{\rm S}+ {\pmb \Pi}_{\rm P} \right),
%\frac{1}{\tau}{\pmb V}_D,\\
\label{tot_momemtum_bal}
\end{split}
\end{align}
where $p$ is the  hydrodynamic pressure, ${\pmb  \Pi}_{\rm S}=\eta_s\left(\nabla {\pmb u}_{\rm S}+(\nabla {\pmb u}_{\rm S})^{\rm T}\right)$ is the Newtonian viscous stress with $\eta_{\rm S}$ being the solvent viscosity, ${\pmb P}_{\rm P}^{\rm osmotic}$ is the additional osmotic stress due to the suspended polymer molecules, and ${\pmb \Pi}_{\rm P}$ is the polymeric elastic stress arising due to the non-local nature of momentum transport via the polymeric back-bone, and as mentioned above, reliant on internal degrees of freedom for its existence. The different stress contributions in the momentum balance above are well understood in terms of their relative importance \cite{bird_vol_1,bird_vol_2}.
\begin{figure}
\subfigure[Bead-centered collision]{
   \includegraphics[scale=0.5]{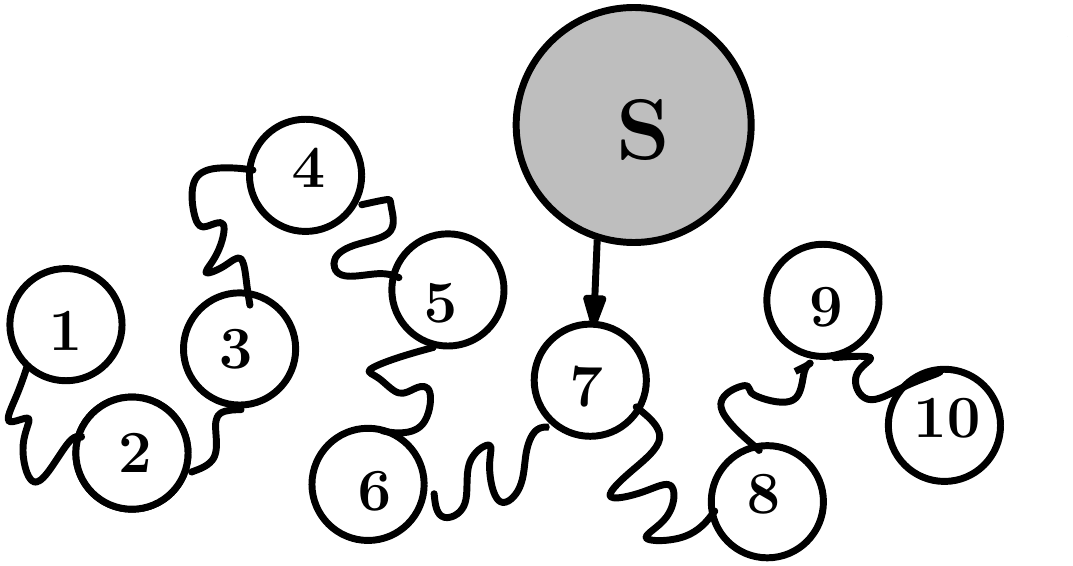}}\qquad
\subfigure[Collision with effective sphere, here ${\rm R_{g}}$ being the radius of gyration]{
\includegraphics[scale=0.5]{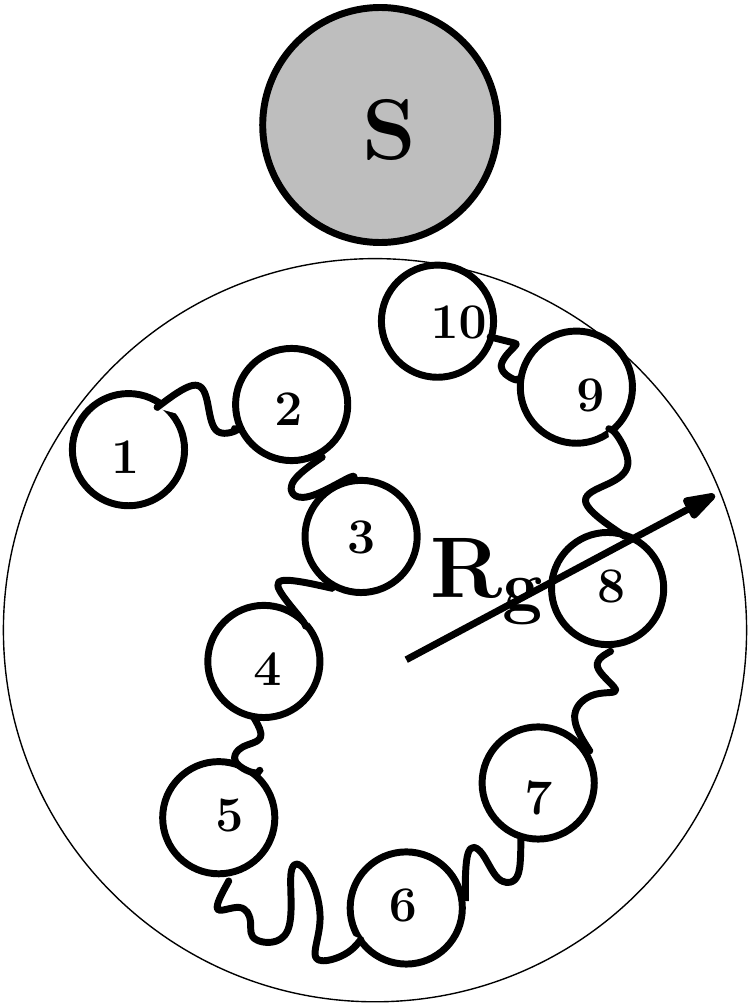}}
\label{cmb}
\caption{Collision mechanisms}
  \end{figure}
The formulation of a phase-space kinetic theory for a polymer-solvent mixture raises the
immediate question as to how to model the  emergence of a  non-local polymeric stress from the local collision picture of  Boltzmann kinetic theory. Unlike the case of a simple gas mixture, such a non-local contribution emerges from describing the polymeric solute (modelled as a bead-spring chain with $N$ beads, say) in terms of an $N$-particle  distribution function. Thus, any detailed kinetic model of a polymeric solution needs to couple the $N$-particle kinetic theory of the solute\,(the precise value of $N$ being dictated by the micro-mechanical model used; $N=2$ for a dumbbell) with the single-particle kinetic theory of the solvent. Such a scenario requires new ingredients to be incorporated in a Boltzmann-type kinetic theory for mixtures of simple gases. For example, what does one mean by a  collisional event?  Does one speak of a collision between a bead and solvent or one between an effective sphere formed by the chain and the solvent molecule (see Fig. \ref{cmb})). What are the collisional invariants and set of slow moments in such a kinetic theory? 
Further, it is not obvious {\it apriori} if, starting from a non-local description of polymer dumbbell, the local collision inherent in Boltzmann kinetic theory can provide a set of slow moments defined in a pointwise manner.
Finally and importantly, how does the well known entropic polymeric stress arise in this kinetic description?  

In the rheological context, the characteristic time scales of interest ensure that the polymer concentration is almost always regarded as uniform. This is reflected in the vast majority of macroscopic constitutive equations in polymer rheology being derived based only on the (internal)\,conformational degrees of freedom of the polymer molecules. The positional degrees of freedom are irrelevant owing to the small center-of-mass diffusivities of the suspended macromolecules, and the resulting long time scales that typically characterize the development of concentration inhomogeneities. There are at least two exceptions to this rule.
The first is the dynamics of polymer-solvent mixtures close to the critical point where the enhanced osmotic compressibility renders concentration fluctuations important. It is known that elastic stresses associated with the dynamics of the inhomogeneous polymer concentration field, when coupled to an ambient shear flow, lead to enhanced scattering in the single phase region above the critical point\cite{Metzner1984,Pine1991}. In attempting to model these concentration fluctuations, which differ qualitatively from those of simple fluid mixtures close to the critical point, researchers have used two-fluid equations at the continuum level\cite{helfand1989large,milner1991hydrodynamics,doi1992dynamic,milner1993dynamical,helfand1994}. In these models, the independent variables  of interest are the polymer and solvent mass and momentum densities. The component mass densities satisfy the respective continuity equations. The momentum balance for the Newtonian solvent involves the familiar viscous stress, while the polymer is also acted on by a combination of osmotic and elastic stresses. In addition, each of these species is acted on by an inter-phase drag force that resists any relative motion. A coupling mechanism of the polymer stress to polymer concentration, as proposed in Ref.\cite{helfand1989large}, is used often to explain the shear banding in polymer solutions \cite{cromer2013shear}. 
The second scenario where the inhomogeneity of the polymer concentration field becomes important is in shearing flows of polymer solutions in confined geometries, specifically microfluidic channels \cite{grahamreview}. In these cases, the polymer residence time becomes long enough to be comparable to the time scale of stress-driven migration in the transverse direction; essentially on account of the disparity between the longitudinal and transverse channel dimensions. There have been several attempts to explain the phenomenon of stress-driven migration that leads to concentration inhomogeneities manifesting as near-wall depletion layers\cite{brunn1,brunn2,magraham2005,grahampabloDNA}. Some of these efforts again are kinetic-theory-based with the solvent still treated as a continuum\cite{Bhave,brunn3}, while others employ a more formal approach based on the Hamiltonian theory of non-equilibrium thermodynamics\cite{Beris1,Beris_taylorcouette}. A third scenario where the diffusive degrees of freedom 
of the suspended microstructure are of importance is shear-banding instabilities that are known to occur in worm-like micellar solutions\cite{Olmsted2008}.

Keeping in mind the aforementioned earlier approaches to the dynamics and rheology of polymer solutions, the two-component phase-space kinetic theory for polymer solutions formulated in this paper leads to a computationally efficient numerical algorithm that  allows for the (1) the characterization of complex flows, both non-viscometric laminar and turbulent, of polymer solutions free of closure approximations that characterize earlier macroscopic constitutive-equation-based approaches (for instance, see \cite{chilcott1988creeping,rallison1988we}); (2) prediction of near-critical dynamics of polymer molecules without the approximation underlying earlier phenomenological descriptions; (3) prediction of stress-driven migration of polymer molecules in confined geometries, and the associated characterization of wall-depletion layers.

The paper is organized as follows. A brief description of the Boltzmann-based kinetic theory of a binary (simple) 
gas mixture is given in section \ref{bin_col}. Then, in section \ref{kin_des},  we describe the kinetic-theory-based approach for a polymer solvent mixture, wherein the polymer is modeled as a  dumbbell and the solvent molecules are structureless particles, and the moment equations for which are consistent with 
the phenomenological description used in the analysis of concentration fluctuations in near-critical polymer solutions. 
In section \ref{col_model_bin}, the collision model for a binary gas mixture 
is discussed. The drawback of the  single relaxation time approximation of BGK collision model is pointed out to begin with, which is that of having a fixed Schmidt (Sc) number. 
This is followed by the introduction of quasi-equilibrium-based collision models with a tunable Schmidt number. Section \ref{collision_model} deals with  a quasi-equilibrium-based 
collision model for a polymer-solvent mixture, which  is shown to reproduce the desired continuum description.
The discrete numerical scheme is discussed in section \ref{num_val} where, starting with the description of the popular two dimensional lattice  model for solvent in section \ref{lb_solvent}, we introduce the  unconventional hyper-lattice  model to solve for the two-particle distribution function of polymer dumbbell  in section \ref{lb_polymer}.
This is followed by a review of the time discretization scheme, and the boundary conditions in the discrete orientation space, in section \ref{discrete_for_4D_1}. 
In section \ref{numerical_validation_kolm}, the effect of polymers in the suppression of inertial instabilities is illustrated for the specific case of a Kolmogorov flow. Finally, the work is summarized in section  \ref{conclusion4d}.

\section{\label{bin_col} The Boltzmann equation for a binary mixture}
\begin{figure}
 \centering \includegraphics[scale=0.5]{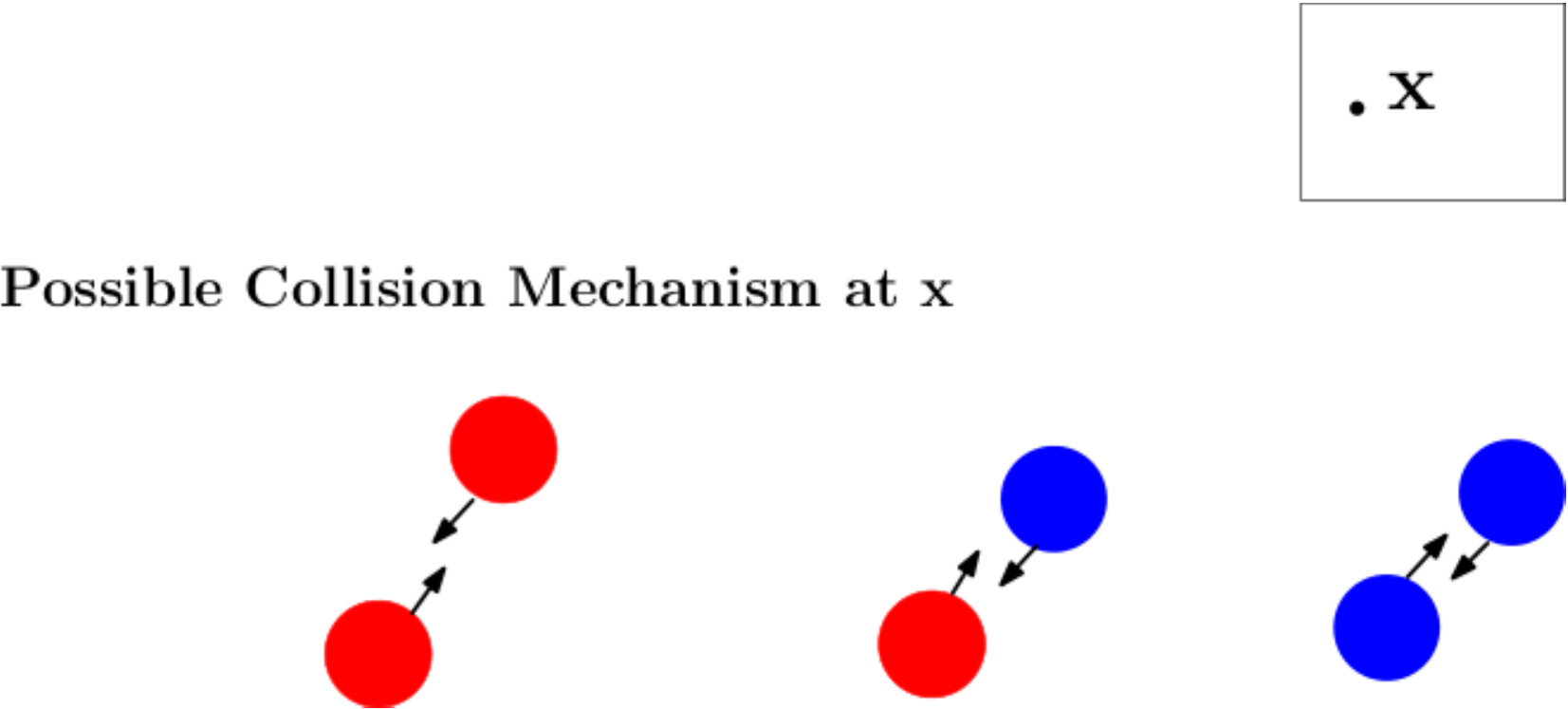}
\caption{Schematic showing different types of collision in a binary gas mixture}
\label{schematic_col_bin}
\end{figure}

In this section, we briefly recall the Boltzmann model as applied to a binary gas mixture \cite{arcidiacono2006simulation,arcidiacono2007simulation}. In a binary gas mixture consisting of two components with masses $m_j$\,($j={\rm A,B}$), in addition to the self-collisions of the A and B particles, cross-collisions between the A and B particles also occur.
Thus, as shown in figure \ref{schematic_col_bin}, three kinds of collisions can occur at a given spatial location ${\pmb x}$ at any instant in time. The kinetic equations governing the evolution of the probability distribution functions of the individual components\,($f_{\rm A}({\pmb x},{\pmb v}_{\rm A},t)$ and $f_{\rm B}({\pmb x},{\pmb v}_{\rm B},t)$) are:
\begin{align}\label{KE:MixtureAB}
\begin{split}
 \frac{\partial }{\partial t}f_{\rm A}({\pmb x},{\pmb v}_{\rm A},t)+{\pmb v}_{\rm A}\cdot\frac{\partial f_{\rm A}}{\partial {\pmb x}} &=\underbrace{\Omega_{AA}(f_{\rm A},f_{\rm A})+\Omega_{AB}(f_{\rm A},f_{\rm B})}_{\Omega_{\rm A}},\\
 \frac{\partial }{\partial t}f_{\rm B}({\pmb x},{\pmb v}_{\rm B},t)+{\pmb v}_{\rm B}\cdot\frac{\partial f_{\rm B}}{\partial {\pmb x}} &=\underbrace{\Omega_{BA}(f_{\rm B},f_{\rm A})+\Omega_{BB}(f_{\rm B},f_{\rm B})}_{\Omega_{\rm B}},
 \end{split}
\end{align}
where, $f_j({\pmb x},{\pmb v}_j,t)$ denotes the probability density of finding a molecule of component $j$\,($j = A$ or $B$) at position ${\pmb x}$ and time $t$. $\Omega_{AA},\Omega_{BB}$ are the self-collision contributions and   $\Omega_{AB}/\Omega_{BA}$ is the  cross-collision contributions which is expressed as \cite{chapman1991mathematical}
\begin{align}
\begin{split}
{\Omega}_{jk}(f_{j},f_{k})&=\int d{\pmb v}_{j}^\prime\,  d{\pmb
v}_{k} \,d{\pmb v}_{k}^\prime \left[ f_{j}({\pmb x},{\pmb v}_{j}^\prime,t)f_{k}({\pmb x},{\pmb v}_{k}^\prime,t)-f_{j}({\pmb x},{\pmb v}_{j},t)f_{k}({\pmb x},{\pmb v}_{k},t) \right]\omega({\pmb v}_{j}^{\prime},{\pmb v}_{k}^{\prime}|{\pmb v}_{j},{\pmb v}_{k}).
\end{split}
\label{bin_col_1}
\end{align} 
The transition probability density, $\omega({\pmb v}_{j}^{\prime},{\pmb v}_{k}^{\prime}|{\pmb v}_{j},{\pmb v}_{k})$ in Eq. \eqref{bin_col_1}, defines the probability that a binary collision between molecules of the components $j$ and $k$ at a given location ${\pmb x}$, with velocities ${\pmb v}_{j}$ and ${\pmb v}_{k}$, leads to velocities ${\pmb v}_{j}^{\prime}$and ${\pmb v}_{k}^{\prime}$ in accordance with the laws of an elastic collision:
\begin{align}
\begin{split}
 m_j {\pmb v}_{j}+  m_k {\pmb v}_{k}&=m_j {\pmb v}_{j}^{\prime}+  m_{k} {\pmb v}_{k}^{\prime}\\
m_j {v}_{j}^2+  m_k {v}_{k}^2&=m_j {v}_{j}^{\prime2}+  m_{k} {v}_{k}^{\prime2}
\label{mom_con}
\end{split}
\end{align}
The transition probability, $\omega$, is symmetric with respect to its dependence on the pre- and post-collisional velocities, that is,
\begin{equation}
 \omega({\pmb v}_{j}^{\prime},{\pmb v}_{k}^{\prime}|{\pmb v}_{j},{\pmb v}_{k})
=\omega({\pmb v}_{j},{\pmb v}_{k}|{\pmb v}_{j}^{\prime},{\pmb v}_{k}^{\prime}).
\end{equation}
reflecting the detailed balance that exists at equilibrium. Self-collisions do not affect mass, momentum and energy conservation. Cross-collisions too do not affect the mass conservation, and one obtains the usual continuity equations for the individual components. However,  momentum and kinetic energy are exchanged between components via cross-collisions in such a manner that the total momentum and energy are conserved. 

Using the kinetic equations \eqref{KE:MixtureAB}, the evolution equations for the component momenta, defined by ${\pmb J}_{j} = \langle m_{j}{\pmb v}_j, f_j \rangle$, are given by \cite{chapman1991mathematical}: 
\begin{align}
\begin{split}
\frac{\partial {\pmb J}_{\rm A}}{\partial t}+ \frac{\partial}{\partial {\pmb x}} \bcdot {\pmb P}_{\rm A}&=\left\langle \Omega_{AB}, m_{\rm A}{\pmb v}_{\rm A}  \right\rangle, \\
\frac{\partial {\pmb J}_{\rm B}}{\partial t}+\frac{\partial}{\partial {\pmb x}} \bcdot {\pmb P}_{\rm B}&= \left\langle \Omega_{BA},  m_{\rm B}{\pmb v}_{\rm B}  \right\rangle.  
\end{split}
\label{ind_mom}
\end{align}
where the component momentum fluxes\,(or stress tensors) are defined by ${\pmb P}_j = \langle m_j {\pmb v}_j{\pmb v}_j, f_j\rangle$ in the above equations, the angular brackets denote a velocity-space average with respect to 
$f_j$, so $<\phi,f_j>=\int f_j\, \phi\,d{\pmb v}_j $.
Using \eqref{bin_col_1}, we get 
\begin{align}
\begin{split}
 \left\langle  \Omega_{AB},m_{\rm A} {\pmb v}_{\rm A}  \right\rangle &=
m_{\rm A}\int  d{\pmb v}_{  A} d{\pmb v}_{  A}^\prime\,  d{\pmb
v}_{ B} \,d{\pmb v}_{  B}^\prime ({\pmb v}_A^\prime-{\pmb v}_A)
f_{  A}({\pmb x},{\pmb v}_{  A}^\prime,t)f_{  B}({\pmb x},{\pmb
v}_{  B}^\prime,t)\omega, \\
 \left\langle  \Omega_{BA}, m_{\rm B} {\pmb v}_{\rm A}  \right\rangle &=
m_{\rm B}\int  d{\pmb v}_{  A} d{\pmb v}_{  A}^\prime\,  d{\pmb
v}_{ B} \,d{\pmb v}_{  B}^\prime ({\pmb v}_B^\prime-{\pmb v}_B)
f_{  A}({\pmb x},{\pmb v}_{  A}^\prime,t)f_{  B}({\pmb x},{\pmb
v}_{  B}^\prime,t)\omega.
\label{col_bin1}
\end{split}
\end{align}
Using momentum conservation given by \eqref{mom_con}, in \eqref{col_bin1}, we get
\begin{equation}
 \left\langle  \Omega_{AB}, m_{\rm A} {\pmb v}_{\rm A}  \right\rangle+\left\langle  \Omega_{BA}, m_{\rm B} {\pmb v}_{\rm B}  \right\rangle=0.
\end{equation}
Thus, the cross-collisions between the two species are solely responsible for momentum exchange, and the corresponding flux can be defined as
\begin{align}
 \begin{split}
{\pmb V}_{\rm D}=\frac{\tau}{2}(\langle\Omega_{AB},m_A{\pmb v}_A\rangle-\langle\Omega_{BA},m_B{\pmb v}_B\rangle).
\label{binary_moments_non_conserved1}
 \end{split}
\end{align}
where ${\pmb V}_{\rm D}$ is the diffusion flux that characterizes the aforementioned exchange process, and the associated time scale $\tau$ is related to the diffusion coefficient  as  $D_{\rm AB}=(X_{\rm A} X_{\rm B}/m_{\rm AB})\tau P $, where $X_j(n_j/n)$ is the individual component mole fraction, $m_{\rm AB} (\rho_{\rm A} \rho_{\rm B}/\rho_{\rm A}+\rho_{\rm B})$ is the reduced mass and $P ( n k_{\rm B} T)$ is the static pressure of the system \cite{arcidiacono2006simulation,arcidiacono2007simulation}. 
The diffusion flux, ${\pmb V}_{\rm D}$, can also be defined in the terms of first order moments in the following form:
\begin{equation}
 {\pmb V}_{\rm D}=m_{\rm AB}\left(\frac{{\pmb J}_{\rm A}}{\rho_{\rm A}}-\frac{{\pmb J}_{\rm B}}{\rho_{\rm B}}\right),
\end{equation}
where $\rho_j =\langle  m_j,f_j\rangle$. The equations for the component momenta, in term of diffusion flux, then take the form
\begin{align}
\begin{split}
\frac{\partial {\pmb J}_{\rm A}}{\partial t}+ \frac{\partial}{\partial {\pmb x}} \bcdot {\pmb P}_{\rm A}&=  \frac{{\pmb V}_{\rm D}}{\tau}, \\
\frac{\partial {\pmb J}_{\rm B}}{\partial t}+\frac{\partial}{\partial {\pmb x}} \bcdot {\pmb P}_{\rm B}&= - \frac{{\pmb V}_{\rm D}}{\tau},
\end{split}
\label{ind_mom}
\end{align}
which are consistent with the total mixture momentum being conserved, shown by 
Eq. \eqref{tot_mom1},
with ${\pmb J} = {\pmb J}_{\rm A} + {\pmb J}_{\rm B}$ and ${\pmb P}={\pmb P}_{\rm A}+{\pmb P}_{\rm B}$.
\begin{align}
\begin{split}
\frac{\partial {\pmb J} }{\partial t}+ \frac{\partial }{\partial {\pmb x}} \bcdot {\pmb P} &=0, \\ 
\end{split}
\label{tot_mom1}
\end{align}.
Similarly, the evolution of the component stress tensors is governed by equations of the form:
\begin{align}
   \begin{split}  
\frac{\partial }{\partial t}{\pmb P}_j  +\frac{\partial }{\partial  {\pmb x}}\cdot 
 {\pmb Q}_{j} 
  &= \langle\Omega_{jj}, m_j{\pmb v}_j {\pmb v}_j\rangle+\langle\Omega_{jk}, m_j{\pmb v}_j {\pmb v}_j\rangle,
\label{binary_moments_mix}
   \end{split}
 \end{align}
where $j\neq k$, and ${\pmb Q}_j$, the flux corresponding to ${\pmb P}_j$, can be written in terms of the distribution function as ${\pmb Q}_j= \langle m_j {\pmb v}_j{ v}_j^2, f_j\rangle$.
The trace of Eq. (\ref{binary_moments_mix}) for $j = A, B$ corresponds to the evolution of the component kinetic energies. Energy conservation implies that total trace is conserved.

The kinetic level description of the Boltzmann type for a binary mixture, as well as the resulting low-order moment equations have been presented above. Here, interactions between the molecules of the two components via cross collisions allow for the exchange of both momentum and kinetic energy, while respecting conversation of the total momentum and kinetic energy. In the next section, based on these considerations, a more elaborate Boltzmann-type description for a polymer-solvent mixture is presented.

\section{\label{kin_des} Extended Boltzmann mixture equation for a polymer solution}
\begin{figure}
 \centering\includegraphics[scale=0.5]{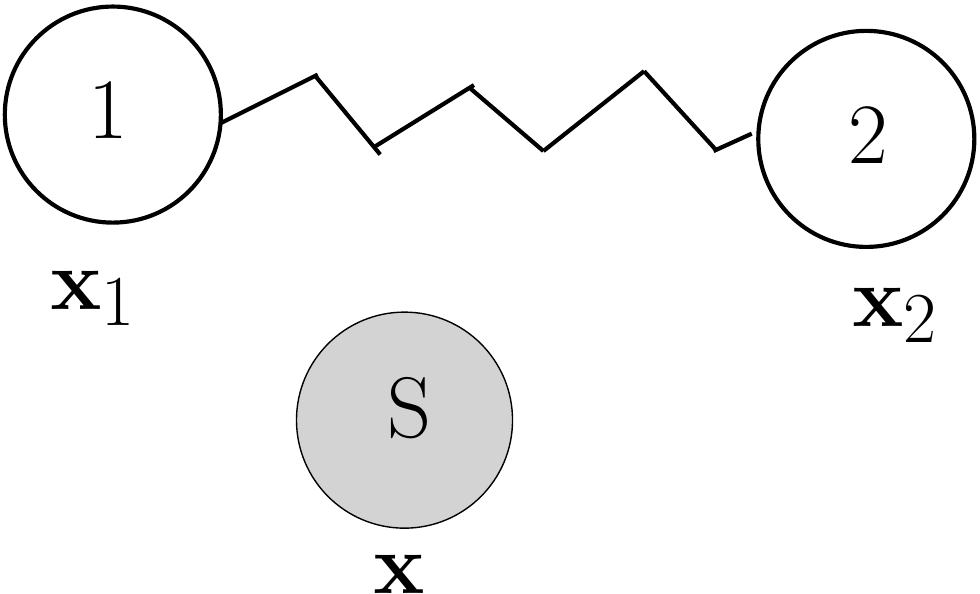}
\caption{Schematic showing the polymer modeled as a dumbbell and solvent as a structure-less particle.}
\label{schematic}
\end{figure}

The simplest micro-mechanical theory of the polymer solution is based on a two component mixture with one of the components being a structureless solvent particle of mass $m_{\rm S}$ and the other component being a polymer dumbbell consisting of two point masses (each of mass $m_{\rm B}$) connected by a massless spring; a schematic of the model appears in Fig. \ref{schematic}. The spring force is a function of the relative separation of the pair of masses, being given by ${\pmb F}_\nu({\pmb x}_\xi-{\pmb x}_\nu)$ (for $\nu,\xi=1,2$) such that ${\pmb F}_1=-{\pmb F}_2$.
Based on the schematic of the model shown in Fig. \ref{schematic}, we extend the Boltzmann paradigm summarized in section \ref{bin_col} to the case of a dilute polymer solutions.

As before, the dynamics of the solvent phase is governed by the single-particle distribution function 
 $ f^{\rm I}_{\rm S}({\pmb x},{\pmb v}_{\rm S},t)$ which denotes the probability of finding a solvent molecule at position ${\pmb x}$ with velocity ${\pmb v}_{\rm S}$ at an instant of time $t$. The subscripts ${\rm S}, {\rm P}$ denote solvent and polymer, respectively; note the added superscript I which helps draw a distinction  to the pair-probability that is relevant to the polymeric dumbbell, and does not appear in the description of the simple gas above. The  solvent mass density $\rho_{\rm S}$, momentum density $\rho_{\rm S} {\pmb u}_{\rm S}$ and temperature $T_{\rm S}$ are defined as:
\begin{align}
\begin{split}
\rho_{\rm S} =\langle  m_{\rm S}, f_{\rm S}^{\rm I} \rangle , \quad
{\pmb J}_{\rm S} =\rho_{\rm S} {\pmb u}_{\rm S} =\langle  m_{\rm S} {\pmb v}_{\rm S},
f_{\rm S}^{\rm I} \rangle,\quad
\rho_{\rm S} T_{\rm S}=\langle m_{\rm S} ({ v}_{\rm S}-u_{\rm S})^2, f_{\rm S}^{\rm I} \rangle.  
\label{sol_moment}
\end{split}
\end{align}
The dynamics of the solute (polymer modeled as a dumbbell) is governed by a two-particle distribution function 
$f_{\rm P}^{\rm II}({\pmb x}_1, {\pmb x}_2, {\pmb v}_{P1}, {\pmb v}_{P2},t)$ which defines the probability of finding the dumbbell such that the bead 1 is at ${\pmb x}_{1}$ with velocity ${\pmb v}_{P1}$ and bead 2 is located at ${\pmb x}_{2}$ with velocity ${\pmb v}_{P2}$ at any instant of time $t$. The mass density of the polymer component at the position ${\pmb x}$ is then defined as:
\begin{align}
\label{Polymerrho}
\begin{split}
 \rho_{\rm P}({\pmb x},t)&= m_{\rm B}\int f_{\rm P}^{\rm II}({\pmb x}_1, {\pmb x}_2,
{\pmb v}_{P1}, {\pmb v}_{P2},t)\delta({\pmb x}-{\pmb x}_1)\,d{\pmb v}_{P1}\,d{\pmb v}_{P2}\,d{\pmb x}_1\,d{\pmb x}_2\\
&+ m_{\rm B}\int f_{\rm P}^{\rm II}({\pmb x}_1, {\pmb x}_2, {\pmb v}_{P1}, {\pmb v}_{P2},t)\delta({\pmb x}-{\pmb x}_2)
\,d{\pmb v}_{P1}\,d{\pmb v}_{P2}\,d{\pmb x}_1\,d{\pmb x}_2,\\
\end{split}
\end{align}
which accounts for contributions of both beads. Therefore, $\rho_{\rm P}=2m_{B}n_{\rm P}$, where $n_{\rm P}$ is the number density of polymers.
Along the same lines, it is natural to define the momentum density and the stress tensor as \citep{ottinger1996kinetic}:
\begin{align}
\begin{split}
\label{Polymermomentum}
{\pmb J}_{\rm P}({\pmb x},t)&=m_{\rm B}\int {\pmb v}_{P1} f_{\rm P}^{\rm II}({\pmb
x}_1, {\pmb x}_2, {\pmb v}_{P1}, {\pmb v}_{P2},t)\delta({\pmb x}-{\pmb x}_1)
\,d{\pmb v}_{P1}\,d{\pmb v}_{P2}\,\,d{\pmb x}_1\,d{\pmb x}_2\\
&+ m_{\rm B}\int {\pmb v}_{P2} f_{\rm P}^{\rm II}({\pmb x}_1, {\pmb x}_2, {\pmb v}_{P1}, {\pmb v}_{P2},t)\delta({\pmb x}-{\pmb x}_2)
\,d{\pmb v}_{P1}\,d{\pmb v}_{P2}\,d{\pmb x}_1\,d{\pmb x}_2.\\
\end{split}\\
\begin{split}
\label{Polymerstress}
{\pmb P}_{{\rm P}}({\pmb x},t)&=m_{\rm B}\int {\pmb v}_{P1}{\pmb v}_{P1}
f_{\rm P}^{\rm II}({\pmb
x}_1, {\pmb x}_2, {\pmb v}_{P1}, {\pmb v}_{P2},t)\delta({\pmb x}-{\pmb x}_1)
\,d{\pmb v}_{P1}\,d{\pmb v}_{P2}\,d{\pmb x}_1\,d{\pmb x}_2\\
&+ m_{\rm B}\int {\pmb v}_{P2}{\pmb v}_{P2} f_{\rm P}^{\rm II}({\pmb x}_1,
{\pmb x}_2, {\pmb v}_{P1},
{\pmb v}_{P2},t)\delta({\pmb x}-{\pmb x}_2)
\,d{\pmb v}_{P1}\,d{\pmb v}_{P2}\,d{\pmb x}_1\,d{\pmb x}_2.\\
 \end{split}
\end{align}
The stress, ${\pmb P}_{\rm P}$ in Eq. \eqref{Polymerstress} only constitutes the  kinetic contribution to the stress tensor, resulting   from the (ballistic) motion of
the beads across a surface. The entropic  stress arising due to  the inter-particle force is discussed later in this section. 
The trace of ${\pmb P}_{\rm P}$ would be the sum of the averaged kinetic energies of the two beads which constitute a part of the osmotic pressure. The total osmotic pressure would be the sum of the kinetic energies of the two beads (compressive) and the trace of the entropic stress (tensile).
Further, as implicit in the definitions above, a solvent-bead collision at the location of interest can occur involve either bead. The momentum balance for each of these collisions may be written as:
\begin{equation}
 m_{\rm S}{\pmb v}_{\rm S}+ m_{\rm B} {\pmb v}_{\rm P1}=m_{\rm S}{\pmb v}_{\rm S1}^{ \prime}+m_{\rm B} {\pmb
v}_{\rm P1}^{ \prime},
\label{at1}
\end{equation}
and
\begin{equation}
 m_{\rm S}{\pmb v}_{\rm S}+ m_{\rm B} {\pmb v}_{\rm P2}=m_{\rm S}{\pmb v}_{\rm S2}^{\prime}+m_{\rm B} {\pmb
v}_{\rm P2}^\prime,
\label{at2}
\end{equation}
respectively. Here, it should be pointed out that the kinetic description of the polymer solution simplifies in terms of a one particle probability distribution defined as
\begin{equation}
 f_{\rm P}^{\rm I}({\pmb x}, {\pmb v}_{\rm P},t)=\int d{\pmb x}_2 d{\pmb v}_{\rm P2}\,f_{\rm
P}^{\rm II}( {\pmb x},{\pmb x}_2,{\pmb v}_{\rm P},{\pmb v}_{\rm P2},t)
+\int d{\pmb x}_1  d{\pmb v}_{\rm P1}\, f_{\rm P}^{\rm II}({\pmb x}_1, {\pmb x}, {\pmb
v}_{\rm P1},{\pmb v}_{\rm P},t),
\label{one_particle}
\end{equation}
which corresponds to the probability of finding either of the beads of the dumbbell at ${\pmb x}$ with   velocity ${\pmb v}_{\rm P}$.
\begin{figure}
 \centering \includegraphics[scale=0.5]{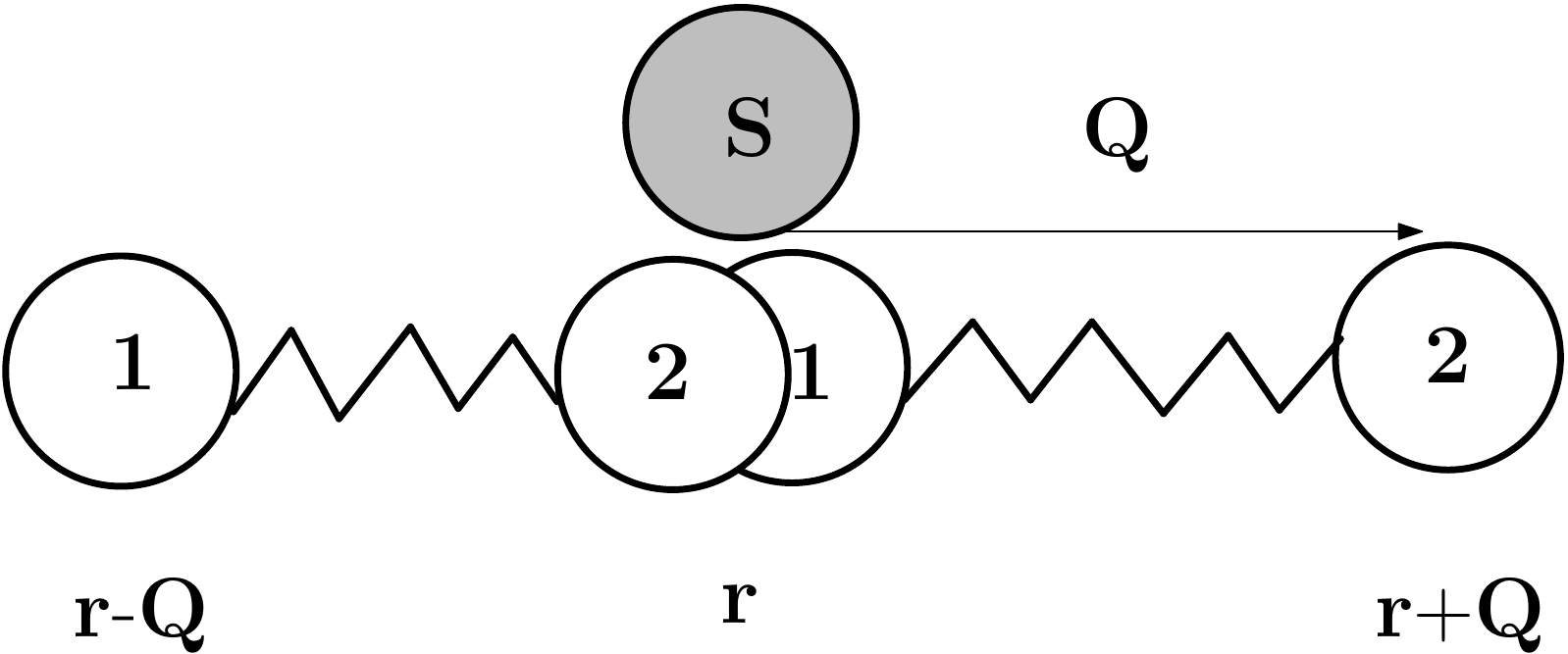}
\caption{Schematic of polymer configuration }
\label{schematic_new}
\end{figure}
Another way of representing the conformation space is shown  Fig. \ref{schematic_new} where the configuration of a polymer dumbbell is defined in terms of ${\pmb r }$ and ${\pmb Q}$ with ${\pmb Q}={\pmb x}_2-{\pmb x}_1$. This corresponds to either bead 1 or 2  being at location ${\pmb r}$ (the other being at ${\pmb r} \pm {\pmb Q}$), with with velocity ${\pmb v}_{\rm P}$. The velocity of the end-to-end vector ${\pmb Q}$ is denoted as $\dot{\pmb Q}$.  The center of mass, in this notation, is located at ${\pmb r}-{\pmb R}_{\nu}$,
where  ${\pmb R}_\nu=(-1)^{\nu}{\pmb Q}/2$ is the vector from the center of mass of the dumbbell to the $\nu^{\rm th}$ bead; the velocity associated with the
center of mass being ${\pmb v}_{\rm P}-\dot{\pmb R}_{\nu}$.
This $({\pmb r},{\pmb Q})$ coordinate system will eventually be used in our kinetic modeling.
The one-particle distribution function for the polymer, as defined by Eq. \eqref{one_particle}, takes the following form in ${\pmb r }-{\pmb Q}$ coordinates.
\begin{equation}
f^{\rm I}_{\rm P}({\pmb r},{\pmb v}_{\rm P},t)=\sum_\nu\int f_{\rm P}^{\rm II}({\pmb r}- 
{\pmb R}_\nu , {\pmb Q}, {\pmb v}_{\rm P}-\dot{\pmb R}_\nu, \dot{\pmb Q},t) \,d{\pmb
Q}\,d\dot{\pmb Q}.
\label{one_fP}
\end{equation}
 
\begin{figure}
\begin{center}
\subfigure[Possibility 1]{
\includegraphics[scale=0.4]{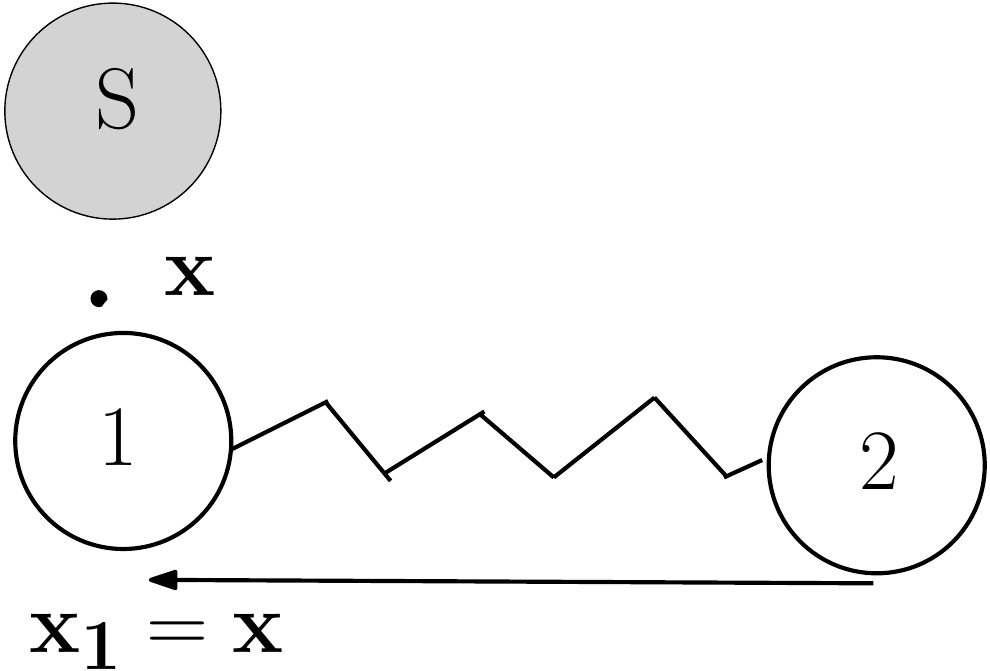}  
}
\subfigure[Possibility 2]{
\includegraphics[scale=0.4]{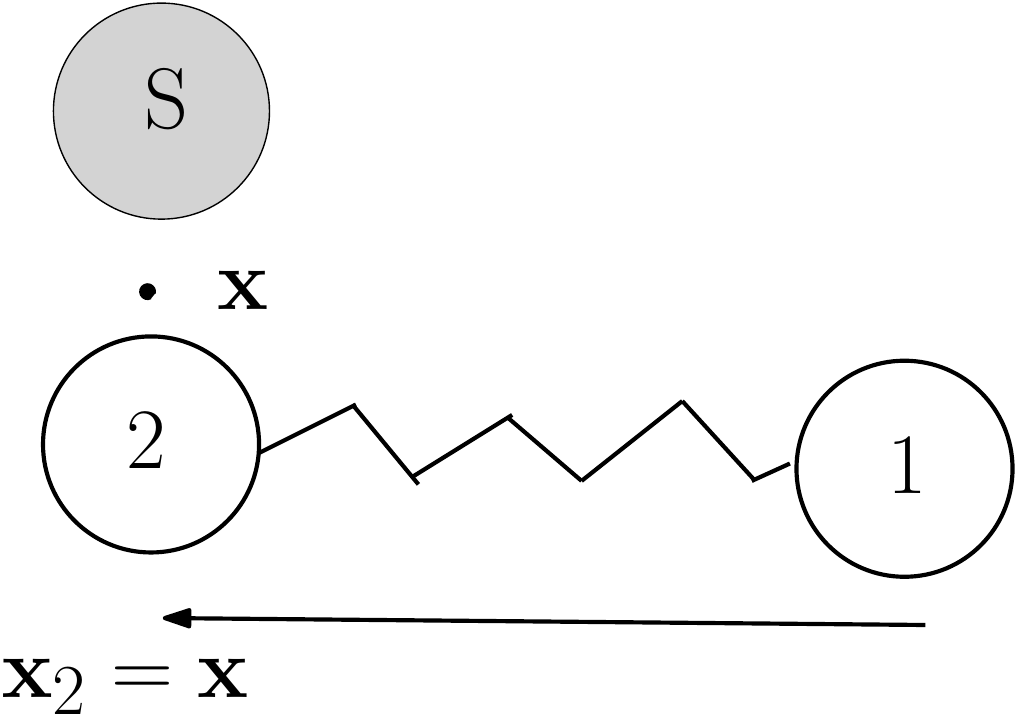}  
} 
\caption {Possible cross collision between solvent molecule and polymer dumbbell at location ${\pmb x}$.}
\label{configuration_poly}
\end{center}
 \end{figure}

The elementary collisions involved in the polymer solution are more complicated owing to the internal degree of freedom associated with the polymer molecule (dumbbell). 
Unlike the binary gas mixture in section \ref{bin_col}, binary cross-collisions are now non-local. Therefore, the polymer dumbbell will collide with the solvent molecule located at ${\pmb x}$ if either of its beads is located at ${\pmb x}$ with the other bead separated by a finite distance ${\pmb Q}$\,(see Fig.\ref{configuration_poly}).  
Having clarified the basic elements involved in the probabilistic description, we now extend the kinetic model of the binary mixture in section \ref{bin_col} to the case of a polymer solution using the collision picture given in Fig.\ref{configuration_poly}. 
The model given below describes the dynamics of the solvent molecules using the one-particle distribution function $f_{\rm S}^{\rm I}({\pmb x},{\pmb v}_{\rm S},t)$ and that of the polymer dumbbells using the two-particle distribution function 
$f_{\rm P}^{\rm II}({\pmb x}_{1}, {\pmb x}_2, {\pmb v}_{P1}, {\pmb v}_{P2},t)$, and in addition, accounts for the non-local collision picture in Fig. \ref{configuration_poly}.

The evolution equation for the solvent probability density, in a manner similar to the simple gas model given in the previous section, can be written as:
 \begin{align}
\begin{split}
 \left(\frac{\partial }{\partial t}+
{\pmb v}_{\rm S}\frac{\partial  }{\partial {\pmb x}} \right)f_{\rm S}^{\rm I}({\pmb x},{\pmb v}_{\rm S},t)
&=  \Omega_{\rm SS}(f_{\rm S}^{\rm I},f_{\rm S}^{\rm I})+ \Omega_{\rm SP}(f_{\rm S}^{\rm I},f_{\rm P}^{\rm II}),
\end{split}
\label{sol_ph}
\end{align}
where $\Omega_{\rm SS}$ accounts for the collision between the solvent molecules, and has a form analogous to the collision terms in section \ref{bin_col} . $\Omega_{\rm SP}$ accounts for the cross-collision between a solvent molecule and a polymer dumbbell, and in explicit form, is given by:
\begin{align}
\begin{split}
 \Omega_{\rm SP}(f_{\rm S}^{\rm I},f_{\rm P}^{\rm II})
&=\int d{\pmb v}_{\rm S1}^\prime\,  d{\pmb
v}_{\rm P1} \,d{\pmb v}_{\rm P1}^\prime\,d{\pmb v}_{\rm P2}^{\prime}
d{\pmb x}_{2} \left[
f_{\rm S}^{\rm I}({\pmb x},{\pmb v}_{\rm S1}^\prime,t)f_{\rm P}^{\rm II}({\pmb x},{\pmb x}_2,{\pmb
v}_{\rm P1}^\prime,{\pmb v}_{\rm P2}^{\prime},t)
-f_{\rm S}^{\rm I}({\pmb x},{\pmb v}_{\rm S},t)
f_{\rm P}^{\rm II}({\pmb x},{\pmb x}_2,{\pmb v}_{\rm P1} ,{\pmb v}_{\rm P2}^{\prime},t)
\right]\omega_1\\
&
+\int d{\pmb v}_{\rm S2}^\prime\,  d{\pmb v}_{\rm P2} \,d{\pmb v}_{\rm P2}^\prime\,d{\pmb
v}_{\rm P1}^{\prime}
d{\pmb x}_{1} \left[
f_{\rm S}^{\rm I}({\pmb x},{\pmb v}_{\rm S2}^\prime,t)f_{\rm P}^{\rm II}({\pmb x}_1,{\pmb x},{\pmb
v}_{\rm P1}^\prime,{\pmb v}_{\rm P2}^{\prime},t)
-f_{\rm S}^{\rm I}({\pmb x},{\pmb v}_{\rm S},t)
f_{\rm P}^{\rm II}({\pmb x}_1,{\pmb x},{\pmb v}_{\rm P1}^\prime ,{\pmb v}_{\rm P2},t)
\right]\omega_2
\end{split} \label{polsolv_coll}
\end{align}
where the following short hand notations
 \begin{align}
 \begin{split}
 \omega_1 \equiv \omega({\pmb v}_{\rm S1}^{\prime},{{\pmb v}_{\rm P1}^{\prime}}|{\pmb
v}_{\rm S},{{\pmb v}_{\rm P1}}), \quad  
 \omega_2 \equiv \omega({\pmb v}_{\rm S2}^{\prime},{{\pmb v}_{\rm P2}^{\prime}}|{\pmb
v}_{\rm S},{{\pmb v}_{\rm P2}}),
 \end{split} 
\end{align}
are used for the transition probabilities.
The first integral  on the right hand side of Eq. \eqref{polsolv_coll} accounts for the
collision happening between a solvent molecule and the first bead of the dumbell. The solvent molecule
moving with velocity ${\pmb v}_{\rm S}$ collides with the first bead moving with velocity ${\pmb v}_{\rm P1}$, and located at ${\pmb x}$, with both switching to post-collisional  velocities ${\pmb v}_{\rm S1}^\prime$ and ${\pmb v}_{\rm P1}^{\prime}$ with probability $\omega_1$. Conversely, a solvent molecule with pre-collisional
velocity ${\pmb v}_{\rm S1}^{\prime}$ can collide with the first bead of the polymer with velocity ${\pmb v}_{\rm P1}^{\prime}$, leading to velocities  ${\pmb v}_{\rm S}$ and ${\pmb v}_{\rm P1}$. Similarly, the second integral term accounts for the collision between solvent and the second bead of the polymer dumbbell.
In terms of the reduced single-particle distribution $f_{\rm P}^{\rm I}$ (see Eq. \eqref{one_particle}), cross-collision term 
may be rewritten as:
\begin{align}
\begin{split}
{\Omega}_{\rm SP}(f_{\rm S}^{\rm I},f_{\rm P}^{\rm I})&=\int d{\pmb v}_{\rm S}^\prime\,  d{\pmb
v}_{\rm P} \,d{\pmb v}_{\rm P}^\prime 
  \left[
f_{\rm S}^{\rm I}({\pmb x},{\pmb v}_{\rm S}^\prime,t)f_{\rm P}^{\rm I}({\pmb x},{\pmb v}_{\rm P}^\prime,t)
-f_{\rm S}^{\rm I}({\pmb x},{\pmb v}_{\rm S},t)f_{\rm P}^{\rm I}({\pmb x},{\pmb v}_{\rm P},t)
\right]\omega_1,
\end{split}
\label{solvent_cross_collsion}
\end{align}
which is now analogous to the cross-collision term in the Boltzmann equation for the simple   
gas mixture as given in Eq. \eqref{bin_col_1}\citep{andries2002consistent}. Similarly, the formal evolution equation for the polymeric solute is:
 \begin{align}
\begin{split}
\left(\frac{\partial }{\partial t}+
{\pmb v}_{\rm P1}\frac{\partial }{\partial {\pmb x}_1}
+{\pmb v}_{\rm P2}\frac{\partial }{\partial {\pmb x}_2}
+\frac{\pmb F_1}{m_{\rm B}}\frac{\partial  }{\partial {\pmb v}_{\rm P1}}
+\frac{\pmb F_2}{m_{\rm B}}\frac{\partial  }{\partial {\pmb v}_{\rm P2}}
\right)f_{\rm P}^{\rm II}({\pmb x}_1,{\pmb x}_2,{\pmb v}_{\rm P1},{\pmb v}_{\rm P2},t)
&=
\Omega_{\rm PS}(f_{\rm S}^{\rm I},f_{\rm P}^{\rm II}), 
 \end{split}
\label{poly_ph}
\end{align}
where ${\pmb F}_1$ and ${\pmb F}_2$ are the spring forces acting on the beads. In this work,  the self-collision between polymer molecules is neglected 
because this contribution is negligibly small in the dilute limit under consideration. The cross-collision term  $\Omega_{\rm PS}$ is given as:  
\begin{align}
\begin{split}
 \Omega_{\rm PS}(f_{\rm_S}^{\rm I},f_{\rm P}^{\rm II})\\
&=\int d{\pmb v}_{\rm S}\,d{\pmb v}_{\rm S1}^\prime\,  d{\pmb v}_{\rm P1}^\prime   
 \left[
f_{\rm S}^{\rm I}({\pmb x}_1,{\pmb v}_{\rm S1}^\prime,t)f_{\rm P}^{\rm II}({\pmb x}_1,{\pmb x}_2,{\pmb v}_{\rm P1}^\prime,{\pmb v}_{\rm P2},t)
-f_{\rm S}(^{\rm I}{\pmb x}_1,{\pmb v}_{\rm S},t)
f_{\rm P}^{\rm II}({\pmb x}_1,{\pmb x}_2,{\pmb v}_{\rm P1} ,{\pmb v}_{\rm P2},t)
\right]\omega_1\\
&+
\int\, d{\pmb v}_{\rm S}\, d{\pmb v}_{\rm S2}^\prime\,  d{\pmb v}_{\rm P2}^\prime\, 
 \left[
f_{\rm S}^{\rm I}({\pmb x}_2,{\pmb v}_{\rm S2}^\prime,t)f_{\rm P}^{\rm II}({\pmb x}_1,{\pmb x}_2,{\pmb v}_{\rm P1}, {\pmb v}_{\rm P2}^{\prime},t)
-f_{\rm S}^{\rm I}({\pmb x}_2,{\pmb v}_{\rm S},t)
f_{\rm P}^{\rm II}({\pmb x}_1,{\pmb x}_2,{\pmb v}_{\rm P1}  ,{\pmb v}_{\rm P2},t)
\right]\omega_2,
\end{split}
\label{polymer_cross_collsion}
\end{align}
where, the first term on right hand side accounts for the collision between a solvent molecule and bead 1  located at ${\pmb x}_1$ and the second term  accounts for the collision between a solvent molecule and bead 2  located at ${\pmb x}_2$. Using the definition of $f_{\rm P}^{\rm I}$ as given in \eqref{one_fP}, \eqref{poly_ph} may again be written in terms of $f_{\rm P}^{\rm I}$ as:
\begin{align}
&\left(\frac{\partial }{\partial t}+
{\pmb v}_{\rm P}\frac{\partial }{\partial {\pmb x}}
\right)f_{\rm P}^{\rm I}({\pmb x},{\pmb v}_{\rm P},t) + \frac{1}{m_{\rm B}}\frac{\partial  }{\partial {\pmb v}_{\rm P}} \left[ \int d{\pmb x}'d{\pmb v}_P^\prime{\pmb F}({\pmb x}-{\pmb x}^\prime)[f_{\rm P}^{\rm II}({\pmb x},{\pmb x}^\prime,{\pmb v}_{\rm P},{\pmb v}_{\rm P}^\prime,t)+ f_{\rm P}^{\rm II}({\pmb x}^\prime,{\pmb x},{\pmb v}_{\rm P}^\prime,{\pmb v}_{\rm P},t)] \right]\nonumber\\
&=\int d{\pmb v}_{\rm S}\,d{\pmb v}_{\rm S}^\prime\,  d{\pmb v}_{\rm P}^\prime   
\left[
f_{\rm S}^{\rm I}({\pmb x},{\pmb v}_{\rm S}^\prime,t)f_{\rm P}^{\rm I}({\pmb x},{\pmb v}_{\rm P}^\prime,t)
-f_{\rm S}^{\rm I}({\pmb x},{\pmb v}_{\rm S},t)
f_{\rm P}^{\rm I}({\pmb x},{\pmb v}_{\rm P},t) \right]\omega_1, \label{singl_fp} 
\end{align}
which bears a closer resemblance to the kinetic equation for the solvent but for the obvious change of subscript\,(${\rm S} \leftrightarrow {\rm P}$). The exception is, of course, the entropic force between the beads that still depends on the pair probability density\,($f_{\rm P}^{\rm II}$).

Apriori it is not obvious that  local conservation laws exist in this system. Therefore, in what follows, the set of conservation laws arising from the kinetic description given by Eq. \eqref{sol_ph} and Eq. \eqref{poly_ph} is discussed. Similar to the Boltzmann equation for the simple gas mixture, cross-collisions conserve mass in the present model. Furthermore, as expected, the total momentum  is conserved, while individual momenta are not; note that, unlike the binary gas mixture, the natural way to define solute momentum density is by Eq. \eqref{Polymermomentum}. 

On integrating \eqref{sol_ph} over all possible values of ${\pmb v}_{\rm S}$, the self collision term goes to zero as before. Using  \eqref{solvent_cross_collsion} for the cross-collision integral term, and the symmetry of the transition probability with respect to pre and post collisional velocities, one gets
\begin{align}
\begin{split}
   \partial_t \rho^{S} + \partial_{\pmb x} \cdot {\bf J}_{\rm S} 
&=m_{\rm S} \int d{\bf v}_{\rm S}   d{\pmb
v}_{\rm S1}^\prime\, d{\pmb
v}_{\rm P1} \,d{\pmb v}_{\rm P1}^\prime
f_{\rm S}^{\rm I}({\pmb x},{\pmb v}_{\rm S1}^\prime )f_{\rm P}^{\rm I}({\pmb x},{\pmb
v}_{\rm P1}^\prime)\omega({\pmb v}_{\rm S1}^{\prime},{{\pmb v}_{\rm P1}^{\prime}}|{\pmb
v}_{\rm S},{{\pmb v}_{\rm P1}})\\
&-m_{\rm S} \int d{\bf v}_{\rm S}   d{\pmb
v}_{\rm S1}^\prime\, d{\pmb
v}_{\rm P1} \,d{\pmb v}_{\rm P1}^\prime f_{\rm S}^{\rm I}({\pmb x},{\pmb v}_{\rm S1}^{\prime})
f_{\rm P}^{\rm I}({\pmb x},{{\pmb v}_{\rm P1}^{\prime}})
 \omega({\pmb
v}_{\rm S},{{\pmb v}_{\rm P1}}|{\pmb v}_{\rm S1}^{\prime},{{\pmb v}_{\rm P1}^{\prime}}),\\
&=0,
\end{split}
\end{align}
which implies the mass conservation for the solvent. Similarly, the evolution of the solvent momentum density of the solvent is given by
\begin{align}
\begin{split}
\partial_t {\pmb J}_{\rm S} + \partial_{\pmb x} \cdot {\pmb P}_{\rm S}  &= 
m_{\rm S}\int \,d{\pmb v}_{\rm S}\, d{\pmb v}_{\rm S1}^\prime\,  d{\pmb
v}_{\rm P1} \,d{\pmb v}_{\rm P1}^\prime 
 \,{\pmb v}_{\rm S} \left[
f_{\rm S}^{\rm I}({\pmb x},{\pmb v}_{\rm S1}^\prime,t)f_{\rm P}^{\rm I}({\pmb x},{\pmb
v}_{\rm P1}^\prime,t)
-f_{\rm S}^{\rm I}({\pmb x},{\pmb v}_{\rm S},t)
f_{\rm P}^{\rm I}({\pmb x},{\pmb v}_{\rm P1},t)
\right]\omega_1\\
&=m_{\rm S}\int  \,d{\pmb v}_{\rm S}\, d{\pmb v}_{\rm S1}^\prime\,  d{\pmb
v}_{\rm P1} \,d{\pmb v}_{\rm P1}^\prime 
 \, \left[{\pmb v}_{\rm S}-{\pmb v}_{\rm S}^\prime\right]f_{\rm S}^{\rm I}({\pmb x},{\pmb
v}_{\rm S1}^\prime,t)f_{\rm P}^{\rm I}({\pmb x},{\pmb
v}_{\rm P1}^\prime,t)\omega_1,
\label{sol_m}
\end{split}
\end{align}
where ${\pmb P}_{\rm S}$ denotes the solvent momentum flux, and similar to the binary gas mixture, the term on the right hand side of the equation accounts for the momentum exchange between the solvent and polymer components.

Unlike the solvent, showing the existence of mass conservation for the polymer phase is a little more subtle owing to the non-locality of the dumbbell.
The evolution equation for the polymer mass density, defined via Eq.\eqref{Polymerrho}, shows the existence of such a conservation law. 
This evolution equation is written, using Eq.\eqref{poly_ph}, as
\begin{align}
\begin{split}
   \partial_t \rho_{\rm P} + \partial_{\pmb x} \cdot {\pmb J}_{\rm P}  &= m_{\rm B}\int d{\pmb
x}_{2} d{\pmb v}_{\rm P1} d{\pmb v}_{\rm P2} 
\Omega_{\rm PS}({\pmb x}_1,{\pmb x}_2,{\pmb v}_{\rm P1},{\pmb v}_{\rm P2},t)\delta({\pmb
x}-{\pmb x}_1)\\
&+m_{\rm B}\int d{\pmb x}_{1} d{\pmb v}_{\rm P1} d{\pmb v}_{\rm P2} 
\Omega_{\rm PS}({\pmb x}_1,{\pmb x}_2,{\pmb v}_{\rm P1},{\pmb v}_{\rm P2},t)\delta({\pmb
x}-{\pmb x}_2),
\end{split}
\end{align}
which, on using symmetry of the transition probability, reduces to the usual continuity equation for the polymer component as
\begin{align}
\begin{split}
   \partial_t \rho_{\rm P} + \partial_{\pmb x} \cdot {\pmb J}_{\rm P}  &= 0,
\end{split}
\end{align}
where the momentum density of the polymer phase ${\pmb J}_{\rm P}$ has been defined in Eq. \eqref{Polymermomentum}.
The evolution equation for the polymer momentum density takes the form:
 \begin{align}
 \begin{split}
 \partial_t {\pmb J}_{\rm P} &+ \partial_{\pmb x} \cdot {\pmb P}_{\rm P} 
 -{\pmb I}\\
=&m_{\rm B}\int d{\pmb v}_{\rm S}\,d{\pmb v}_{\rm S1}^\prime\,  d{\pmb v}_{\rm P1}^\prime\,
d{\pmb v}_{\rm P1} \, d{\pmb v}_{\rm P2}  \,d{\pmb x}_2
{\pmb v}_{\rm P1} \left[
 f_{\rm S}^{\rm I}({\pmb x},{\pmb v}_{\rm S1}^\prime)f_{\rm P}^{\rm II}({\pmb x},{\pmb
x}_2,{\pmb v}_{\rm P1}^\prime,{\pmb v}_{\rm P2})  
- f_{\rm S}^{\rm I}({\pmb x},{\pmb v}_{\rm S})
f_{\rm P}^{\rm II}({\pmb x},{\pmb x}_2,{\pmb v}_{\rm P1} ,{\pmb v}_{\rm P2})  
\right]\omega_1\\
+&
 m_{\rm B} \int\, d{\pmb v}_{\rm S}\, d{\pmb v}_{\rm S2}^\prime\,  d{\pmb v}_{\rm P2}^\prime\,
d{\pmb v}_{\rm P1} \,  d{\pmb v}_{\rm P2}
\,  d{\pmb x}_{1}
 {\pmb v}_{\rm P2} \left[
 f_{\rm S}^{\rm I}({\pmb x},{\pmb v}_{\rm S2}^\prime)f_{\rm P}^{\rm II}({\pmb x}_1,{\pmb
x},{\pmb v}_{\rm P1}, {\pmb v}_{\rm P2}^{\prime}) 
- f_{\rm S}^{\rm I}({\pmb x},{\pmb v}_{\rm S},t)
f_{\rm P}^{\rm II}({\pmb x}_1,{\pmb x},{\pmb v}_{\rm P1}  ,{\pmb v}_{\rm P2}) 
\right]\omega_2\, \\
=&m_{\rm B} \int d{\pmb v}_{\rm S}\, d{\pmb v}_{\rm S1}^\prime\,  d{\pmb
v}_{\rm P1} \,d{\pmb v}_{\rm P1}^\prime 
 \,  \left[{\pmb v}_{\rm P}-{\pmb v}_{\rm P}^\prime \right]
f_{\rm S}^{\rm I}({\pmb x},{\pmb
v}_{\rm S1}^\prime)f_{\rm P}^{\rm I}({\pmb x},{\pmb
v}_{\rm P1}^\prime)\omega_1, 
 \end{split} \label{polyph_mom}
\end{align}
where the symmetry of the transition probability has again been used for the collision term. The term ${\pmb I}$ on the left hand side of Eq. (\ref{polyph_mom}) is defined as:
\begin{equation}
{\pmb I}({\pmb x},t)=  \int   
F({\pmb x}_2-{\pmb x})\psi({\pmb x},{\pmb x}_2,t)d{\pmb x}_2 
- \int F({\pmb x}-{\pmb x}_1)\psi({\pmb x}_1,{\pmb x},t)d{\pmb x}_1,
\label{impulse}
\end{equation}
where the condition ${\pmb F}_1=-{\pmb F}_2\equiv{\pmb F}$ is used, with the configuration distribution function $\psi$ being defined as:
\begin{equation}
 \psi({\pmb x}_1,{\pmb x}_2,t)=\int d{\pmb v}_1\,d{\pmb v}_2 f^{\rm II}_{\rm P}({\pmb
x}_1,{\pmb x}_2,{\pmb v}_1,{\pmb v}_2,t).
\end{equation}
%${\pmb I}$ denotes the non-local momentum transfer due to the stretching of the spring.
The local collision of the solvent molecule with individual bead will result in an impulse  which is communicated down the backbone of the polymer dumbbell. This can also be understood as the  non-local momentum transfer due to stretching of the polymer spring, the effect
of which in polymer momentum density evolution (Eq. \eqref{polyph_mom}) is represented by the term ${\pmb I}$.
The integral of ${\pmb I}$ over all space is given by
\begin{equation}
 \int d{\pmb x} \, {\pmb I}({\pmb x},t)=\int d{\pmb x} d{\pmb x}_2 F({\pmb
x}_2-{\pmb x})\psi({\pmb x},{\pmb x}_2,t) 
-  \int d{\pmb x}d{\pmb x}_1 F({\pmb x}-{\pmb x}_1)\psi({\pmb x}_1,{\pmb x},t)
=0,
\end{equation}
Thus, global momentum conservation is not affected by ${\pmb I}$, and it can, in fact, be defined as the divergence of a second order tensor as:
\begin{equation}
 {\pmb I}= \frac{\partial }{\partial {\pmb r}}\cdot{\pmb 
\Theta}.
\label{IDef}
\end{equation}
To see this, we note that  ${\pmb I}$ (Eq. \eqref{impulse}), can be re-written in $({\pmb r},{\pmb Q})$ coordinates as 
\begin{equation}
{\pmb I}({\pmb r},t)= \sum_\nu \int  {{\pmb F}_{\nu}({\pmb Q})} 
\psi({\pmb r}- {\pmb R}_\nu , {\pmb Q},t) d{\pmb Q},
\end{equation} 
Further, assuming the configuration probability density to vary slowly over a dumbbell length, and expanding the configuration distribution function $\psi$  in a Taylor series 
(\cite{ottinger1996kinetic}) as
 \begin{equation}
 {\psi}({\pmb r}-{\pmb R}_{\nu}, {\pmb Q}, t)= {\psi}({\pmb r}, {\pmb Q}, t)
-{\pmb R}_{\nu}\cdot\frac{\partial }{\partial {\pmb r}} {\psi}({\pmb r}, {\pmb
Q}, t)
+\frac{{\pmb R}_{\nu}{\pmb R}_{\nu}}{2}{\pmb :}\frac{\partial }{\partial {\pmb
r}}\frac{\partial }
{\partial {\pmb r}} {\psi}({\pmb r}, {\pmb Q}, t)
+...,
\end{equation}
which gives \eqref{IDef} with 
\begin{align}
 { \pmb  \Theta}({\pmb r},t)=&\int  {\psi}({\pmb r},{\pmb Q}, t)\,{  \pmb Q}\, {
 \pmb F}\,d{\pmb Q},
 %\\
%=& \langle {\pmb F}{\pmb Q} \rangle,
\end{align}
which is the usual form of the polymeric configurational stress tensor; for Hookean dumbbells, the expression reduces to the spring constant $H$ times the conformation tensor given as $\int{\psi}({\pmb r},{\pmb Q}, t) {\pmb Q}{\pmb Q} \,d{\pmb Q}$.
Equation (\ref{polyph_mom}) 
therefore takes the form
 \begin{align}
	\begin{split}
		\partial_t {\pmb J}_{\rm P} &+ \partial_{\pmb x} \cdot {\pmb P}_{\rm P} 
		-\partial_{\pmb x}\cdot{\pmb \Theta}\\
		=&m_{\rm B} \int d{\pmb v}_{\rm S}\, d{\pmb v}_{\rm S1}^\prime\,  d{\pmb
			v}_{\rm P1} \,d{\pmb v}_{\rm P1}^\prime 
		\,  \left[{\pmb v}_{\rm P}-{\pmb v}_{\rm P}^\prime \right]
		f_{\rm S}({\pmb x},{\pmb
			v}_{\rm S1}^\prime)f_{\rm P}^{\rm I}({\pmb x},{\pmb
			v}_{\rm P1}^\prime)\omega_1, 
	\end{split} \label{polyph_mom2}
\end{align}
It should be noted that while the above expansion of the configuration probability density, in yielding the usual elastic stress tensor, 
is restricted to the characteristic flow dimension being much larger than the polymer radius of gyration, the kinetic theory formulation above is not limited by this assumption, and in principle, allows for a non-local
stress tensor in cases where the flow or geometric dimension starts to become comparable to the radius of gyration \cite{brunn1985kinetic}.
Finally, the evolution equation for the total momentum density ${\pmb J}={\pmb J}_{\rm P} +{\pmb J}_{\rm S}$, obtained by adding those for the component momenta (Eq.\eqref{sol_m} and Eq.\eqref{polyph_mom2}) is   
\begin{align}
 \begin{split}
 \frac{\partial {\pmb J}}{\partial t}     + 
%\frac{\partial {p}_{\rm P}}{\partial  {\pmb r}} 
\frac{\partial  }{\partial{ {\pmb  r}}}\cdot \left({{\pmb P}}_{\rm P}+{{\pmb P}}_{\rm S}
-{\pmb \Theta}\right)
 &=  0,
\label{total_momentum}
\end{split}
\end{align}
This conservation form for the total momentum density also implies that the evolution of the momentum densities of the solvent and polymer (Eqs. \eqref{sol_ph},\eqref{poly_ph}) can, similar to the gas mixture, be re-written in terms of a diffusion velocity as:
 \begin{align}
 \begin{split}
\frac{\partial {\pmb J}_{\rm S} }{\partial t} +
%\frac{\partial {p}_{\rm S}}{\partial  {\pmb r}}+
\frac{\partial }{\partial  {\pmb r}}\cdot{\pmb P}_{\rm S}({\pmb r},t) 
&=\frac{1}{\tau}{\pmb V}_{\rm D}\\
 \frac{\partial {\pmb J}_{\rm P}}{\partial t}     
%\frac{\partial {p}_{\rm P}}{\partial  {\pmb r}} 
+ \frac{\partial  }{\partial{ {\pmb  r}}}\cdot{{\pmb P}_{\rm P}}({\pmb r}, t) 
 &=-\frac{1}{\tau}{\pmb V}_{\rm D}
+ \frac{\partial }{\partial {\pmb r}}\cdot{\pmb  \Theta}
,\\
\label{poly_sol_mom1}
\end{split}
\end{align}
where, using  Eq. \eqref{sol_m}, ${\pmb V}_{\rm D}$ is defined as:
\begin{align}
 {\pmb V}_{\rm D}=& \tau\, m_{\rm S}\int  \,d{\pmb v}_{\rm S}\, d{\pmb
v}_{\rm S1}^\prime\,  d{\pmb
v}_{\rm P1} \,d{\pmb v}_{\rm P1}^\prime 
 \, \left[{\pmb v}_{\rm S}-{\pmb v}_{\rm S}^\prime\right]f_{\rm S}^{\rm I}({\pmb x},{\pmb
v}_{\rm S1}^\prime,t)f_{\rm P}^{\rm I}({\pmb x},{\pmb
v}_{\rm P1}^\prime,t)\omega_1, \\
=-& \tau\, m_{\rm B}\int  \,d{\pmb v}_{\rm S}\, d{\pmb
v}_{\rm S1}^\prime\,  d{\pmb
v}_{\rm P1} \,d{\pmb v}_{\rm P1}^\prime 
 \, \left[{\pmb v}_{\rm P}-{\pmb v}_{\rm P}^\prime\right]f_{\rm S}^{\rm I}({\pmb x},{\pmb
v}_{\rm S1}^\prime,t)f_{\rm P}^{\rm I}({\pmb x},{\pmb
v}_{\rm P1}^\prime,t)\omega_1.
\end{align}
Here, $\tau $ can again be understood as a time scale associated with the  drag force which resists the velocity difference between the two components \citep{milner1991hydrodynamics,milner1993dynamical}. 
%to remain nearly equal 
% {\bf \color{red}GS: Need to have a sentence after this to emphasize that this is the same system of equations solved by Fredrickson, Milner and others (references mentioned in the introduction).}
To conclude, in this section starting from a Boltzmann-like kinetic
description of the solvent-polymer mixture in phase space, 
a set of conversation laws, analogous to those obtained in Refs \cite{helfand1989large,milner1991hydrodynamics,doi1992dynamic,milner1993dynamical,helfand1994}, have been obtained for the polymer solution. Indeed, these equations
must be reproduced by any model equation written for this system. In subsequent
sections, a simple BGK-type model is developed, where these equations are used as
consistency conditions.   

\section{\label{col_model_bin}Collision model for binary gas mixture}

Having introduced the kinetic theory framework for both the binary gas and the polymer-solvent mixtures, we now move on to a brief description of the corresponding collision models for purposes of numerical implementation. As already seen in section \ref{bin_col}, any self-consistent collision model for the binary gas mixture should obey the  following properties:
\begin{itemize}
\item The self-collision does not affect mass, momentum and energy conservation.
\begin{equation}
  \left\langle  \Omega_{jj},  m_j\left\lbrace\begin{aligned}
     1 \\
      {\pmb v}_j \\
      \frac{v_j^2}{2}
    \end{aligned}\right\rbrace\right\rangle=0. 
\end{equation}
\item The cross-collision does not affect mass conservation, but leads to momentum and energy exchanges between components such that the total momentum and energy are conserved.
\begin{align}
 \left\langle m_j \Omega_{jk}\right\rangle=&\,0 \,\,{\rm with }\,\,j\neq k\,(=A,B), \\
   \begin{aligned}\left\langle \Omega_{AB},
      m_A  {\pmb v}_A  \right\rangle +\left\langle  \Omega_{BA},
      m_B {\pmb v}_B  \right\rangle     
    \end{aligned} =&\,0, \\
    \begin{aligned}\left\langle \Omega_{AB},
      m_A  \frac{{\pmb v}_A^2}{2}  \right\rangle +\left\langle  \Omega_{BA},
      m_B\frac{{\pmb v}_B^2}{2}  \right\rangle     
    \end{aligned} =&\,0, \label{cross-collision}
\end{align}
with $j = A, B$.
\item   Indifferentiability: the mixture  description reduces  to the single component description 
when the components become mechanically equivalent. Thus, when $m_{\rm A}=m_{\rm B}$, the total distribution $f=f_A+f_B$, must obey the single species Boltzmann equation \citep{andries2002consistent,sirovich1}.

\item Similar to the original Boltzmann equation, the collision model should also have an $H$- theorem of the form
\begin{equation}
 \frac{\partial H}{\partial t} +\frac{\partial }{\partial {\pmb r}}\cdot {\pmb J}_{\rm H}= -\sigma,
\end{equation}
with $\sigma\geq0$. Here, the  $H$ function is defined as 
\begin{equation}
H=\sum_{j}^{  A,B }\int m_j f_j(\log {f_j}-1) d{\pmb v}, 
\end{equation}
with the flux of $H$-function given by
\begin{equation}
 {\pmb J}_{\rm H}=\sum_{j}^{  A,B }\int m_j f_j(\log {f_j}-1){\pmb v}_j\, d{\pmb v}_j,
\end{equation}
and the entropy production being given by
\begin{equation}
 \sigma=\sum_{j}^{A,B}\langle m_j\log{f_j},\Omega_j\rangle.
\end{equation}
Furthermore, the entropy production ${\pmb \sigma }=0$ if and only if $ f_j=f_j^{\rm MB}(M^{\rm Slow})$
which  implies
\begin{equation}
 \Omega_j=0 \iff f_j=f_j^{\rm MB}(M^{\rm Slow}).
\end{equation}
 \end{itemize}
where $f_j^{MB}$ refers to the local Maxwell-Boltzmann distribution for the $j^{\rm th}$ component, and $M^{\rm Slow}$ refers to the slow manifold comprising the appropriate hydrodynamic variables  \cite{cercignani1988boltzmann,succi_book}.  

In what follows, we first describe in brief the BGK and quasi-equilibrium approximations for the collision operator, as applied to a binary gas mixture; the following section deals with the quasi-equilibrium models for the polymer-solvent mixture. One of the simplest and most widely used models for the collision operator is
the single-relaxation time approximation, known as a Bhatnagar-Gross-Krook (BGK)
approximation \cite{bhatnagar1954model}. Herein, the collision kernel, $\Omega_j=\Omega_{jj}+\Omega_{jk}$, is defined as \citep{andries2002consistent}
\begin{equation}
 \Omega_j= \frac{1}{\tau}(f_j^{\rm MB}(\rho_j, {\pmb U}, T)-f_j),
\end{equation}
where   ${\pmb U} = {\pmb J}/\rho$ is the total mixture velocity  and $\rho =\rho_A + \rho_B$ is the mixture mass density. This gives
 the following form for the  rate of change of the non-conserved mixture moments 
%Eq.\eqref{binary_moments_non_conserved} will then become
\begin{align}
 \begin{split}
\frac{1}{2}(\langle\Omega_A,m_A{\pmb v}_A\rangle-\langle\Omega_B,m_B{\pmb v}_B\rangle)
= -\frac{1}{\tau}{\pmb V}_{\rm D}, \quad
 \sum_{j}^{\rm A,B}\langle\Omega_j,m_j{\pmb v}_j{\pmb v}_j\rangle
=-\frac{1}{\tau}({\pmb P} - {\pmb P}^{\rm eq}),
\label{binary_moments_non_conserved1}
 \end{split}
\end{align}
 where, ${\pmb P}^{\rm eq} = nk_BT_0 {\pmb I} + \frac{{\pmb J}{\pmb J}}{\rho}$.
 Equation \eqref{binary_moments_non_conserved1} shows that for the BGK model, the mass  diffusion flux and the pressure tensor relax on the same time scale $\tau$, which results in a fixed Schmidt number, $Sc$ (the ratio of the momentum and mass diffusivities) of order unity.
 One needs at least two different time scales associated with the relaxation rates of the mass diffusion and momentum fluxes, which suggests that the usual BGK collision kernel 
is not an appropriate model for binary gas mixtures.
The single relaxation time  approximation is even more inappropriate for polymer-solvent
mixtures where due to low center-of-mass diffusivities, polymer mass transfer modes  have the extremely long relaxation times, in turn leading to very large values of Sc.

In Refs. \cite{arcidiacono2006simulation,arcidiacono2007simulation,ansumali2007quasi}, a collision model for binary mixtures, based on an intermediate quasi-equilibrium state, has been proposed in order to have a tunable ${\rm Sc}$. 
\begin{figure}
\begin{center}
\includegraphics[scale=0.5]{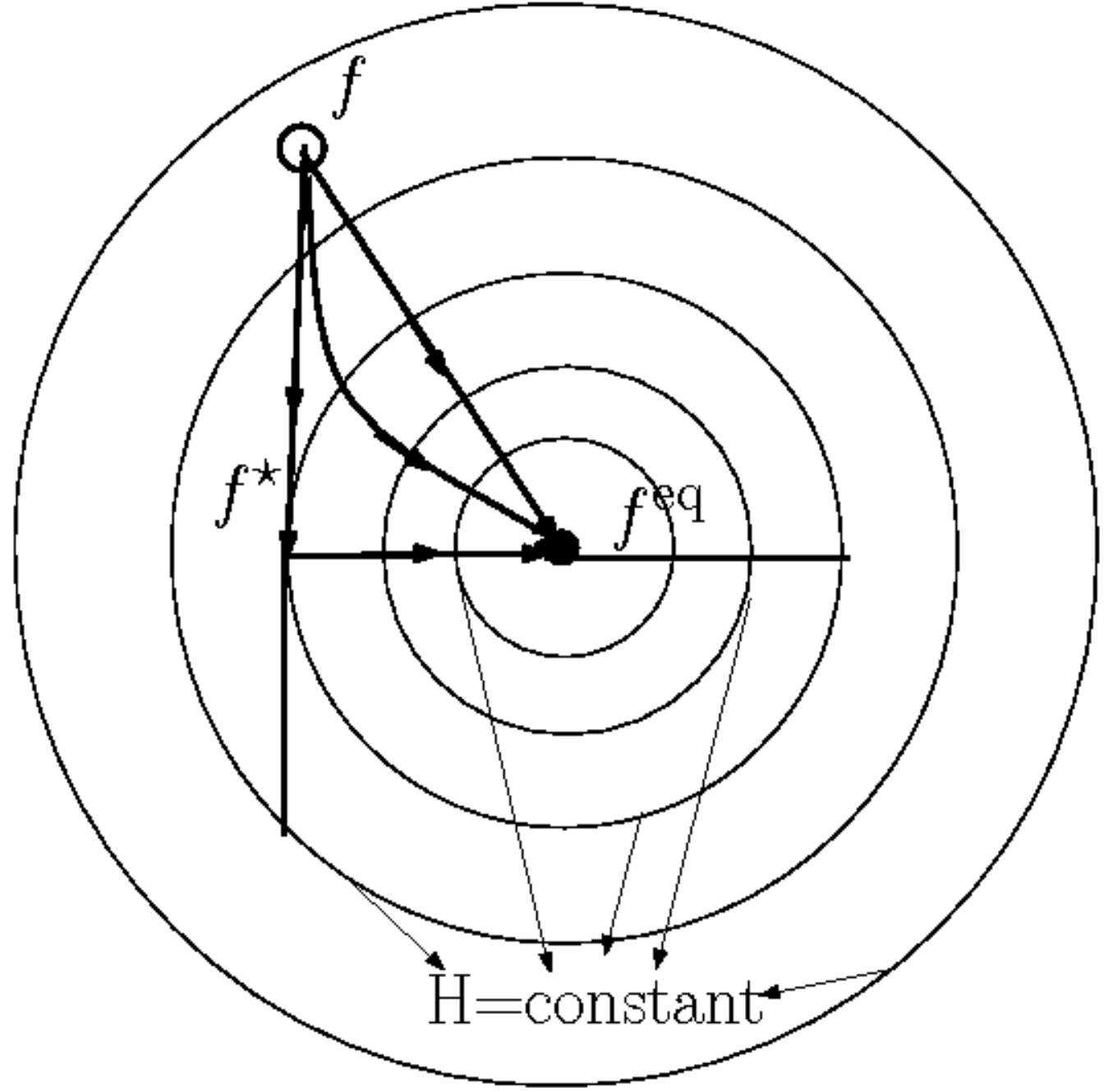}
 \caption{  Scheme showing the relaxation of $f$ to $f^{\rm eq}$ through a quasi-equilibrium state $f^{\star}$.}
\label{quasi_equli_fig}
\end{center}
\end{figure}
They followed the concept of a quasi-equilibrium as explained in Fig  \ref{quasi_equli_fig}.
%The concept of a quasi-equilibrium is explained in Fig. \ref{quasi_equli_fig}. 
As shown therein, there is a fast relaxation of the distribution function $f$
towards the quasi-equilibrium $f^{\star}$, followed by a slow relaxation towards the
equilibrium state $f^{\rm eq}$. Both stages of relaxation can be modeled as BGK-type terms with $\tau_1^{-1}$ and $\tau_2^{-1}$ as the respective rates of relaxation.
The equilibrium distribution function $f^{\rm eq}$ is evaluated in the usual manner by minimizing the $H$-function under the constraints of fixed slow variables $M^{\rm Slow}$. The quasi-equilibrium, $f^{\star}$, is found by the minimizing the $H$-function  under the constraints of fixed  quasi-slow variables which, in the present case, can be taken as the individual component momenta\,\cite{arcidiacono2006simulation} or the stresses\,\cite{ansumali2007quasi}.
The simplest generalization of the BGK model using $f^{\star}$ and the individual component momenta as quasi-conserved variable can be written as:
\begin{equation}
 \Omega_j=\frac{1}{\tau_1}[f^{ \star}_j(\rho_j,{\pmb u}_j, T_j)-f_j]+\frac{1}{\tau_2}
[f^{ \rm eq}_j(\rho_j,{\pmb U}, T)-f^{\star}_j(\rho_j,{\pmb u}_j, T_j)].
\label{quasieq1}
\end{equation}
where the component velocities are defined by ${\pmb u}_j = {\pmb J}_j/\rho_j$. It is worth noting that in order to satisfy the $H$-theorem, a proper ordering of the relaxations is required which in the present case corresponds to $\tau_1\leq\tau_2$ \citep{gorban_karlin_quasi_eq,ansumali2007quasi}. Using the fact ${\pmb P}^{\star}={\pmb P}^{\rm eq}$, it can be  seen that that 
\begin{align}
 \begin{split}
\sum_{j}^{\rm A,B}\langle\Omega_j,m_j{\pmb v}_j{\pmb v}_j\rangle
=-\frac{1}{\tau_1} ({\pmb P} - {\pmb P}^{\rm eq}), \quad 
\frac{1}{2}(\langle\Omega_A,m_A{\pmb v}_A\rangle-\langle\Omega_B,m_B{\pmb v}_B\rangle)
= -\frac{1}{\tau_2} {\pmb V}_{\rm D},
\label{binary_moments_non_conserved2}
 \end{split}
\end{align} 
so that the pressure tensor and the diffusion mass flux now relax on different time scales.
A Chapman-Enskog expansion shows that the first order non-equilibrium contributions to the  pressure tensor ${\pmb P}$ and the mass diffusion flux ${\pmb V}_{\rm D}$\,(note that ${\pmb V}^{\rm eq}_{\rm D} = 0$) 
are \citep{arcidiacono2006simulation}
\begin{align}
 \begin{split}
 {\pmb P}^{\rm neq}&={\pmb P} - {\pmb P}^{\rm eq}\\
&=-\tau_1 n k_{\rm B} T\left[\left(\frac{\partial  {\pmb U}}{\partial {\pmb r}} \right)+
\left(\frac{\partial {\pmb U}}{\partial {\pmb r}} \right)^{\rm T}
-\frac{2 {\pmb \delta}}{D}\frac{\partial }{\partial {\pmb r}}\cdot{\pmb U}
\right], \\
{\pmb V}_{\rm D}&=\tau_2 k_{\rm B} \left[ \frac{\rho_A}{\rho}\frac{\partial (n_{B} T )}{\partial {\pmb r}}
- \frac{\rho_B}{\rho}\frac{\partial( n_{A} T) }{\partial {\pmb r}}\right].
\label{noneq_contri}
\end{split}
 \end{align}
%, it can be shown that the  viscosity coefficient $\mu$ is related to relaxation 
%time $\tau_1$ as $\mu=nk_{\rm B} T \tau_1$,

It is evident from  Eq. \eqref{noneq_contri} that the shear viscosity  $\mu$ is proportional to the relaxation time $\tau_1$ as $\mu=nk_{\rm B} T \tau_1$. Further, and after some rearrangement, the  diffusion coefficient $D_{AB}$ can be  related to the  relaxation 
time $\tau_2$ giving tunable ${\rm Sc}$ where ${\rm Sc}=\mu/(\rho D_{AB})$. Although tunable, ${\rm Sc}$ is not arbitrary. The choice of the  quasi-equilibrium defined by \eqref{quasieq1}, and the implied ordering of the relaxation times, leads to an upper bound on $ {\rm Sc}$:
$ {\rm Sc}\leq {\rm Sc}^{\star}$. The  threshold Schmidt number ${\rm Sc}^\star$ depends on the component mass fraction $Y_j(\rho_j/\rho)$ and mole fractions\,$X_j(n_j/n)$, being given by $ {\rm Sc}^{\star}=(Y_A Y_B)/(X_A X_B)$; the details of the calculation can be found in \cite{arcidiacono2006simulation}. If the component molecular masses, $m_j$, are of the same order, 
${\rm Sc}^{\star}$ comes out to be the ratio of masses in  the dilute limit, and thus use of \eqref{quasieq1} restricts one to ${\rm Sc}'s$ of order unity or smaller. This is a particularly severe limitation for the polymer-solvent system of  interest since, as already mentioned, the small diffusivities of the polymer molecules imply that the typical Schmidt numbers for such systems are very large.

In order to avoid the aforementioned ${\rm Sc}$ limitation, the elements of the stress tensor  ${\pmb P}_j$  of individual components, can instead be chosen as the set of quasi-conserved variables, with
the  slow variables being the  individual mass densities $\rho_j$ and total momentum density ${\pmb J} = \rho\,{\pmb U}$, for purposes of minimizing the $H$-function . Denoting the resulting quasi-equilibrium as $f^{\star\star}(\rho_j, {\pmb U}, {\pmb P}_j)$,
the collision integral takes the following form:
\begin{equation}
 \Omega_j=\frac{1}{\tau_1}[f^{ \star\star}_j(\rho_j,{\pmb U}, {\pmb P}_j)-f_j]+\frac{1}{\tau_2}
[f^{ \rm eq}_j(\rho_j,{\pmb U}, T)-f^{\star\star}_j(\rho_j,{\pmb U}, {\pmb P}_j)].
\label{quasieq2}
\end{equation}
%{\bf \color{red} GS: Mention of the fact that the quasi-equilibria in this case have to be found numerically.}
The non-conserved mixture moments now take the form:
\begin{align}
 \begin{split}
\sum_{j}^{\rm A,B}\langle\Omega_j,m_j{\pmb v}_j{\pmb v}_j\rangle
=-\frac{1}{\tau_2} {\pmb P}^{\rm neq}, \quad 
\frac{1}{2}(\langle\Omega_A,m_A{\pmb v}_A\rangle-\langle\Omega_B,m_B{\pmb v}_B\rangle)
= -\frac{1}{\tau_1} {\pmb V}_{\rm D}^{\rm neq},
\label{binary_moments_non_conserved3}
 \end{split}
\end{align} 
and a Chapman-Enskog expansion, similar to the above case, leads to the the  expressions for the  pressure tensor and the mass diffusion flux same as given by Eq. \eqref{noneq_contri} but with the only difference that $\tau_1$ and $\tau_2$ are interchanged . This means that the viscosity $\mu$ is now related to $\tau_2$ and  the  diffusion coefficient $D_{AB}$ to $\tau_1$
The limitation on ${\rm Sc}$ is therefore reversed, being given by ${\rm Sc}\geq{\rm Sc}^{\star}$, which is appropriate to the polymer-solvent mixture.
Thus, between them, the two (component momenta and stress-tensor based) quasi-equilibria formulations cover the entire range of $\rm Sc$ \cite{arcidiacono2006entropic,arcidiacono2006simulation,arcidiacono2007simulation}.

\section{\label{collision_model}Collision Modeling for polymer-solvent mixture}
As discussed in section \ref{kin_des}, the polymer dumbbell collides with the solvent molecule
only if the location of the solvent coincides with the location of either of the beads of dumbbell.
In order to properly handle the non-local polymer-solvent interaction, the required
%Thus, any collision model of polymer-solvent interaction has to be non-local. 
%Therefore, the equation for solvent 
%Recall from  that the collision model for this system has to be developed for 
system of kinetic equations are  given by:
 \begin{align}
\begin{split}
\left(\frac{\partial }{\partial t}+ {\pmb v}_{\rm S}\cdot\frac{\partial   }{\partial {\pmb r}}\right)f_{\rm S}^{\rm I}({\pmb r}, {\pmb v}_{\rm S},t) &= 
\Omega_{\rm S} = \Omega_{\rm SS} + \Omega_{\rm SP},  \\
\left(\frac{\partial }{\partial t}+
{\pmb v}_{\rm 1}\cdot\frac{\partial }{\partial {\pmb x}_{\rm 1}}
+ {\pmb v}_{\rm 2}\cdot\frac{\partial }{\partial {\pmb x}_{\rm 2}} + \frac{{\pmb F}_1}{m_B}\cdot\frac{\partial }{\partial {\pmb v}_{\rm 1}} + \frac{{\pmb F}_2}{m_B}\cdot\frac{\partial }{\partial \dot{\pmb x}_2} \right)f_{\rm P}^{\rm II}({\pmb x}_{\rm 1},{\pmb x}_{\rm 2},{\pmb v}_{\rm 1},{\pmb v}_{\rm 2},t)
&=
 \Omega_{\rm PS},
%}
 \end{split}
\label{solute_ph}
\end{align}
where the collision operators $\Omega_{\rm S}$ and $\Omega_{\rm PS}$ should be modeled such that the continuum level description,  given by  \eqref{poly_sol_mom1}, is recovered. Similar to the mixture model for the binary gas, one needs two relaxation times in order to have a tunable ${\rm Sc}$, and in particular, to be able to access the large ${\rm Sc}$'s of interest.

On using the quasi-equilibrium-based relaxation method described above, with the component momenta being the quasi-conserved variables, the solvent collision term in \eqref{solute_ph} takes the form:
\begin{align}
\begin{split}
 \Omega_{\rm S}&=\frac{1}{\tau_1}\left[f_{\rm S}^{\rm MB}(\rho_{\rm S},{\pmb u}_{\rm
S}, T_{\rm S})-f_{\rm S}^{\rm I}\right]
+\frac{1}{\tau_2}\left[f^{\rm MB}_{\rm S}(\rho_{\rm S}, {\pmb U}, T)-
f_{\rm }^{\rm MB}(\rho_{\rm S},{\pmb u}_{\rm S}, T_{\rm S})\right],\\
\end{split}
\end{align}
where $f_{\rm S}^{\rm MB}(\rho_{\rm S}, {\pmb u}_{\rm S}, T)$ is the Maxwell-Boltzmann distribution about solvent velocity ${\pmb u}_{\rm S}$ and  $f_{\rm S}^{\rm MB}(\rho_{\rm S}, {\pmb U}, T)$ is the Maxwell-Boltzmann distribution about the solution velocity ${\pmb U}$. 
The collision term in the polymer kinetic equation must account for the collisions with each of the two beads of the dumbbell; recall that, in $({\pmb r},{\pmb Q})$ coordinates, the bead coordinates corresponding to these collisions are $({\pmb x}_{\rm 1},{\pmb x}_{\rm 2}) \equiv ({\pmb r},{\pmb r}+{\pmb Q})$ and $({\pmb x}_{\rm 1},{\pmb x}_{\rm 2}) \equiv ({\pmb r}-{\pmb Q},{\pmb r})$; the corresponding coordinates for the center-of-mass and configuration\,(the dumbbell end-to-end vector) are $({\pmb r}+\frac{\pmb Q}{2},{\pmb Q})$ and $({\pmb r}-\frac{\pmb Q}{2},{\pmb Q})$, respectively. Thus, one may write:
\begin{align}
\Omega_{\rm PS}(f_{\rm S}^{\rm I},f_{\rm P}^{\rm II}) \equiv \Omega_{\rm PS}^{(1)}[f_{\rm S}^{\rm I},f_{\rm P}^{\rm II}({\pmb r}+\frac{\pmb Q}{2},{\pmb Q},{\pmb v}_{\rm P}+\frac{\dot{\pmb Q}}{2},\dot{\pmb Q},t)] + \Omega_{\rm PS}^{(2)}[f_{\rm S}^{\rm I},f_{\rm P}^{\rm II}({\pmb r}-\frac{\pmb Q}{2},{\pmb Q},{\pmb v}_{\rm P}-\frac{\dot{\pmb Q}}{2},\dot{\pmb Q},t)].
\end{align}
where each of the $\Omega^{(\nu)}_{\rm PS}$'s are given by a quasi-equilibrium ansatz similar to that of the solvent above:
\begin{align}
\Omega^{(\nu)}_{\rm PS}=\frac{1}{\tau_1}[f^{\star II}_{\rm P}  -f_{\rm P}^{\rm II} ]+\frac{1}{\tau_2}
[ f^{{\rm eq}II}_{\rm P} 
-f^{\star II}_{\rm P}  ],
\label{fp_x1_x2}
\end{align}
with the arguments of the distributions involved being different for $\nu = 1$ and $2$. 
%{\bf \color{red} GS: Is there a reason you have changed the superscript from i to $\nu$. This change is not reflected in the text.}
%{\color{blue}I have changed it to $\nu$ simply because earlier we have used $\nu$ to define two configration of cross-collision between polymer and solvent.}
Thus, the equilibrium distributions in $\Omega^{(1)}_{\rm PS}$ and $\Omega^{(2)}_{\rm PS}$ are
\begin{align}
 \begin{split}
f^{\rm eq II}_{\rm P}({\pmb r}+\frac{\pmb Q}{2},{\pmb Q},{\pmb v}_{\rm P}+\frac{\dot{\pmb Q}}{2},\dot{\pmb Q})
&=\psi({\pmb r}+\frac{\pmb Q}{2},{\pmb Q}) \left(\frac{m^{\rm
B}}{2\pi
k_{\rm B} T}\right)^3
\times\\
&\exp{\left[-\left(\frac{m^{\rm B}({\pmb v}_{\rm P}-{\pmb U}({\pmb
r})-\frac{{\pmb F}_1}{\zeta})^2}{2k_{\rm B} T}\right)-
\left(\frac{m^{\rm B}({\pmb v}_{\rm P}+\dot{\pmb Q}-{\pmb U}( {\pmb r}+{\pmb
Q})-\frac{{\pmb F}_2}{\zeta})^2}{2k_{\rm B} T}\right)\right]},\\
f^{\rm eq II}_{\rm P}({\pmb r}-\frac{\pmb Q}{2},{\pmb Q},{\pmb v}_{\rm P}-\frac{\dot{\pmb Q}}{2},\dot{\pmb Q})
&=\psi({\pmb r}-\frac{\pmb Q}{2},{\pmb Q})  \left(\frac{m^{\rm B}}{2\pi k_{\rm B} T}\right)^3\times \\
&\exp{\left[-
	\left(\frac{m^{\rm B}({\pmb v}_{\rm P}-{\pmb U}( {\pmb r})-\frac{{\pmb F}_2}{\zeta})^2}{2k_{\rm B} T}\right)-\left(\frac{m^{\rm B}({\pmb v}_{\rm P}-
\dot{\pmb Q}-{\pmb U}({\pmb r}-{\pmb Q})-\frac{{\pmb F}_1}{\zeta})^2}{2k_{\rm B} T}\right)\right]},
\label{feq_eq_poly}
\end{split}
\end{align}
respectively, and the corresponding quasi-equilibria are
\begin{align}
 \begin{split}
f^{\star II}_{\rm P}({\pmb r}+\frac{\pmb Q}{2},{\pmb Q},{\pmb v}_{\rm P}+\frac{\dot{\pmb Q}}{2},\dot{\pmb Q})
&=\psi({\pmb r}+\frac{\pmb Q}{2},{\pmb Q}) \left(\frac{m^{\rm
B}}{2\pi k_{\rm B} T}\right)^3\times\\
&\exp{\left[-\left(\frac{m^{\rm B}[{\pmb v}_P-{\pmb u}^{\rm r}({\pmb r}+\frac{\pmb Q}{2}]^2)}{2k_{\rm B} T}\right)- \left(\frac{m^{\rm B}[{\pmb v}_P +\dot{\pmb Q}- ({\pmb u}^{\rm r}({\pmb r}+\frac{\pmb Q}{2})
+{\pmb u}^{\rm Q}({\pmb r}+\frac{\pmb Q}{2} ))]^2}{2k_{\rm B} T}\right)\right]},
\\
f^{\star II}_{\rm P}({\pmb r}-\frac{\pmb Q}{2},{\pmb Q},{\pmb v}_{\rm P}-\frac{\dot{\pmb Q}}{2},\dot{\pmb Q}) &=\psi({\pmb r}-\frac{\pmb Q}{2},{\pmb Q}) \left(\frac{m^{\rm B}}{2\pi k_{\rm B} T}\right)^3
\times\\
&\exp{\left[-
	\left(\frac{m^{\rm B}[{\pmb v}-{\pmb u}^{\rm r}({\pmb r}-\frac{\pmb Q}{2})]^2}{2k_{\rm B} T}\right)
	- \left(\frac{m^{\rm B}[{\pmb v}_P -\dot{\pmb Q}-({\pmb u}^{\rm r}({\pmb r}-\frac{\pmb Q}{2} )-{\pmb u}^{\rm Q}({\pmb r}-\frac{\pmb Q}{2},{\pmb Q}))]^2}{2k_{\rm B} T}\right)\right]}
.
\label{quasi_eq_poly}
\end{split}
\end{align}
Note that the $f_{\rm P}^{\rm eq II}$ and $f_{\rm P}^{ \star II}$ are factorized Maxwellians in ${\pmb r}$ and ${\pmb Q}$ space with $\psi$ corresponding to the pair probability characterizing the dumbbell configuration.
The velocities used in equilibrium distributions (Eq. \ref{feq_eq_poly}) comes from the local velocity of the solution, ${\pmb U}$, whereas the velocities in quasi equilibrium distribution (Eq. \ref{quasi_eq_poly}) are the local velocity of the polymer phase given as $ \psi\,{\pmb u}^{\rm r}=\int \, {\pmb v}_{\rm P} f_{\rm P}^{\rm II}\,d{\pmb v}_{\rm P}\,d\dot{\pmb Q}$ and $\psi\,{\pmb u}^{\rm Q}=\int\,\dot{\pmb Q}\, f_{\rm P}^{\rm II} d{\pmb v}_{\rm P}d\dot{\pmb Q}$.
As already discussed, one requirement of the above model is that it should recover the continuum description involving the spatial coordinate\,(${\pmb r}$) alone, detailed in section \ref{kin_des}, after integration over the remaining degrees of freedom. A further, stricter, requirement is that the Smoluchowski equation for the configuration distribution function in $({\pmb r},{\pmb Q})$ space must be recovered from the primitive phase-space description, given by \eqref{solute_ph} and \eqref{fp_x1_x2}, after integration over the velocity degrees of freedom. In order to show that the model does lead to the expected form of the Smoluchowski equation over longer length and time scales, we first define bead averaged version of any quantity $\phi$ in configuration space as $\hat{\phi}({\pmb r}, {\pmb Q}, t)=\sum_{\nu}\phi({\pmb r}-{\pmb R}_{\nu}, {\pmb Q}, t)$.
 Using this definition, the evolution equation
for the lower order moments for Eq. \eqref{fp_x1_x2} takes the following form
\begin{align}
 \begin{split}
 \frac{\partial }{\partial t}  {\psi}  + &  
\frac{\partial  }{\partial{\pmb   r }} \cdot
 {\pmb J}^{\rm r}   
+ \frac{\partial  }{\partial{\pmb   Q}} \cdot
 {\pmb J}^{\rm Q}   
=0, \\
 \frac{\partial }{\partial t}  {\pmb J}^{\rm r, Q}      + & 
 \frac{\partial  }{\partial{ {\pmb  r}}}\cdot {\pmb P}^{\rm r, rQ}  
+ \frac{\partial  }{\partial{  {\pmb Q}}}\cdot {\pmb P}^{\rm rQ,Q}  
=\frac{1}{\tau_2}\left[{\pmb J}^{\rm  r, Q}_{eq}
 - {\pmb J}^{\rm r,Q}  \right],
\label{polymer_moment_sum}
\end{split}
\end{align}
where   ${\pmb J}^r({\pmb r}, {\pmb Q}, t)$ and ${\pmb J}^Q({\pmb r}, {\pmb Q}, t)$ are the phase-space averaged momentum density for ${\pmb v}_{\rm P}$ and $\dot{\pmb Q}$ respectively. In other words, 
	${\pmb J}^r = \langle\langle {\pmb v}_{\rm P}\rangle\rangle$ and ${\pmb J}^Q  =\langle\langle  \dot{\pmb Q}\rangle\rangle$ with the operator $\langle\langle.. \rangle\rangle$ for any arbitrary quantity $\phi$ being defined as $\langle\langle\phi\rangle\rangle =\sum_\nu\int d{\pmb v}_{\rm P}d\dot{\pmb Q} \phi f_{\rm P}^{\rm II}({\pmb r}+{\pmb R}_{\nu},{\pmb Q},{\pmb v}_{\rm P}+{\dot{\pmb Q}}_{\nu},\dot{\pmb Q})$ with $\sum_{\nu}$ describing the sum over the contribution of both the beads. Similarly, ${\pmb P}^{\rm r}$, ${\pmb P}^{\rm Q}$
	and ${\pmb P}^{\rm rQ}$ are the phase-space averaged second order stress tensors represented as  $\langle\langle {\pmb v}_{\rm P}{\pmb v}_{\rm P}\rangle\rangle$, $\langle\langle {\pmb Q}\dot{\pmb Q}\rangle\rangle$ and $\langle\langle  {\pmb v}_{\rm P}\dot{\pmb Q}\rangle\rangle$ respectively.

	{The explicit form of quasi-equilibria distribution function  (Eq. \eqref{quasi_eq_poly}) results in 
	$\langle\langle   ({\pmb v}_{\rm P},\dot{Q})\rangle\rangle_{\star}$  to be ${\pmb J}^{\rm r, Q}$, hence cancelling the contribution of first terms of collsion operator as represented in Eq. \eqref{fp_x1_x2}. Here the subscript of $\lambda$ ($\lambda$ being $\star$ or eq) on the operator $\langle\langle..\rangle\rangle$ defines the distribution function $f^{\rm \lambda II }$.
	with respect to which averages are taken. 	
	The time scale, $ \tau_2$, is now associated with momentum relaxation process since $\langle\langle   ({\pmb v}_{\rm P},\dot{Q})\rangle\rangle_{\rm eq}$ takes the following form
	\begin{align}
 \begin{split}
  {\pmb J}^{\rm  r}_{eq}&={\psi}{\pmb U} ({\pmb r},t)  
 + \sum_\nu\left(\frac{{\pmb F}_\nu}{\zeta}\psi({\pmb r}-{\pmb R}_{\nu}, {\pmb Q}, t) \right),\\
  {\pmb J}^{\rm  Q}_{eq}&= \psi({\pmb r}, {\pmb Q}, t) 
  {\pmb Q}\cdot\frac{\partial {\pmb U}}{\partial {\pmb r}}- {\psi}
 \frac{ 2{\pmb F} }{\zeta}.
 \label{poly_imposed_vel}
 \end{split}
\end{align}
	
	 }
	 At this point, it is worth mentioning that a  Chapman-Enskog expansion (as  detailed in Appendix \ref{appen}), shows that the dynamics at the O(1) is the desired Smoluchowski equation which governs the evolution of  ${\psi}$ in conformation (${\pmb r}-{\pmb Q}$) space and is given as
 \begin{align}
 \begin{split}
 \frac{\partial \psi}{\partial t}    + &  
\frac{\partial  }{\partial{\pmb   r } }  \left\{
\psi{ \pmb U} +\sum_\nu\left(\frac{{  \pmb F}_{\nu}}{\zeta}\psi({\pmb r}-{\pmb R}_{\nu}, {\pmb Q}, t) \right) \right\}
+ \frac{\partial  }{\partial{  \pmb  Q}}  \left(
 \psi {\pmb Q}\cdot\frac{\partial {\pmb U}}{\partial {\pmb r}}-\psi \frac{2{\pmb F}}{\zeta}\right)
 =\frac{k_{\rm B}T}{\zeta}\left(\frac{\partial^2\psi}{\partial {\pmb r}^2}+2\frac{\partial^2\psi}{\partial {\pmb Q}^2}\right).
\end{split}
\label{Smoluchowski_new11}
\end{align}
In the dilute limit, the above equation recovers the desired Smoluchowski
equation for the homogeneous case \cite{laso1993calculation,feigl1995connffessit} and diffusion equation (for the polymer concentration) for the inhomogeneous case \cite{beris1994compatibility,ottinger1996kinetic,apostolakis2002stress} (see Appendix). 
By integrating out the  conformation degrees of freedom,  the  polymer mass density $\rho_{\rm P}$ , the momentum density ${\pmb J}_{\rm P}$, and stress tensor given by Eq. \eqref{Polymerrho}, Eq. \eqref{Polymermomentum} and Eq. \eqref{Polymerstress}, can also be defined in the following manner 
\begin{align}
 \begin{split}
  \rho_{\rm P}({\pmb r}, t)=\int d{\pmb  Q}\,\,   m_{\rm B}{\psi}({\pmb r},{\pmb Q}, t),\quad
{\pmb J}_{\rm P}({\pmb r}, t)=\int d{\pmb  Q}\,\,   m_{\rm B}{\pmb J}^{\rm r}({\pmb r},{\pmb Q}, t), \quad
{\pmb P}_{\rm P}({\pmb r}, t)=\int d{\pmb  Q}\,\,   m_{\rm B}{\pmb P}^{\rm r}({\pmb r},{\pmb Q}, t).
 \end{split}
\end{align}
 Subsequently, the equations \eqref{polymer_moment_sum} together with solvent description gives the  individual mass conservation represented as 
 \begin{align}
 \begin{split}
  \frac{\partial \rho_{\rm (S,P)}}{\partial t}+\frac{\partial {\pmb J}_{\rm (S,P)}}{\partial {\pmb r}}&=0,\\
 \end{split}
\end{align} 
 and, momentum conservation as 
  \begin{align}
 \begin{split}
  %\frac{\partial \rho_{\rm (S,P)}}{\partial t}+\frac{\partial {\pmb J}_{\rm (S,P)}}{\partial {\pmb r}}&=0,\\
 %\frac{\partial \rho_P}{\partial t}+\frac{\partial \rho_{\rm P}}{\partial {\pmb r}}&=0,\\
 \frac{\partial {\pmb J}_{\rm S}}{\partial t}+\frac{\partial {\pmb P}_{\rm S}}{\partial {\pmb r}}&=\frac{{\pmb V}_{\rm D}}{\tau_2},\\
 \frac{\partial {\pmb J}_{\rm P}}{\partial t}+\frac{\partial }{\partial {\pmb r}}({\pmb P}_{\rm P}-{\pmb \Theta})&=-\frac{{\pmb V}_{\rm D}}{\tau_2}.
 \end{split}
\end{align} 
where the solvent and polymer phase exchange momentum through the drag term $ {\pmb V}_{\rm D}$. It should be emphasized that these 
are the set of continuum equation which are desired from the present kinetic model \cite{helfand1989large,milner1991hydrodynamics,doi1992dynamic,milner1993dynamical,helfand1994}.
%are the exact equation which  set continuum equation any present kinetic model is de
The drawback of this model is that it will limit the maximum attainable ${\rm Sc}$ to be equal to mass ratio in the limit of dilute solution \cite{arcidiacono2006simulation,ansumali2007quasi}. 
 In order to avoid this limitation, the relevant collision model  is \cite{ansumali2007quasi}
\begin{align}
\begin{split}
 \Omega_{\rm S}&=\frac{1}{\tau_1}\left[f_{\rm S}^{\star }(\rho_{\rm S},{\pmb U}, {\pmb P}_{\rm S})-f_{\rm S}\right]
+\frac{1}{\tau_2}\left[f^{\rm MB}_{\rm S}(\rho_{\rm S}, {\pmb U})-
f_{\rm S}^{\star}(\rho_{\rm S},{\pmb U},{\pmb P}_{\rm S})\right],\\
 \Omega_{\rm P}&=\frac{1}{\tau_1}\left[f_{\rm P}^{\star \star II}(\psi,{\pmb U}, {\pmb P}^{{\rm r, rQ, Q}}) -f^{\rm II}_{\rm P}\right]
+\frac{1}{\tau_2}\left [f^{{\rm eq} II}_{\rm P}(\psi,{\pmb U}) -
f_{\rm P}^{\star \star II}(\psi,{\pmb U}, {\pmb P}^{{\rm r, rQ, Q}}) \right].
\end{split}
\end{align}
 such that  $\langle f_{\rm S}^{\star },{ \pmb  v}_{S}{  \pmb v}_{S}\rangle=
 {  \pmb P}_{S }$ and $\langle \langle ({ \pmb  v}_{P}{  \pmb v}_{P},{ \pmb  v}_{P}\dot{  \pmb Q},\dot{  \pmb Q}\dot{  \pmb Q})\rangle\rangle_{\star \star}={\pmb P}^{{\rm r, rQ, Q}}$.

This model will give  the moment-chain  same as Eq. \eqref{polymer_moment_sum} 
 but with  the relaxation time $\tau_1$ instead 
of  $\tau_2$ and therefore the lower limit on ${\rm Sc}$ will become ${\rm Sc}^\star$ 
for dilute  solution,   which     was the  upper limit in the previous model.
Physically, the two models  differ in terms of the fixed quasi  variables. 
In the first model where ${\rm Sc}^\star$ is the upper limit, 
the  velocity of individual component is a quasi variable.
It 
means that the system first relaxes to a state with a fixed  component velocity 
and then relaxes to a state which   has fixed  mass averaged velocity.
In the model where  ${\rm Sc}^\star$ is the lower limit, the quasi  
variable is the pressure tensor of the individual component.

\section{Numerical Scheme\label{num_val}} 
The lattice Boltzmann is conventionally used as  Navier-Stokes equations solver.  In recent years, we have shown that the diffusive dynamics of momentum relaxation of  polymer molecule which is otherwise  governed by Fokker Planck equation can effectively be recovered using a BGK type relaxation \cite{singh2011lattice,singh2013lattice}. In this section, using a two dimensional set-up, we develop a discrete two fluid kinetic model
for polymer  based on LB mechanism. The framework used to represent the solvent and polymer phase are discussed in in subsections \ref{lb_solvent} and \ref{lb_polymer} respectively. % 
% In this section, we propose a discrete  two fluid kinetic model for polymer solution based on LB mechanism.  Using a two dimensional set-up, we describe the lattice Boltzmann  framework used to represent the scheme discussed above for solvent and polymer respectively, in subsections \ref{lb_solvent} and \ref{lb_polymer}.
In the LB formulation,  one  works  with a set of discrete populations $f=\{f_i\}$ which corresponds to predefined discrete velocities   ${\bf c}_i$ $(i=1,\cdots,N)$ to represent the original continuous system \cite{benzi1992lattice,succi_book}.
 \subsection{\label{lb_solvent}Lattice Boltzmann  model  for solvent}
 The solvent phase is represented by probability distribution function $f_{S}$ ( superscript `${\rm I}$' is removed for simplicity) and the  discrete evolution  equation of interest is
%  To represent the solvent, we need to solve the  following discrete equation of evolution of probability distribution function $f_{Si}$() :
 \begin{equation}
  \partial_t f_{Si} + c_{i\alpha} \partial_{\alpha} f_{Si} = \frac{1}{\tau_1}\left(f^{ \star  }_{Si}  
-f_{Si}^{\rm  }\right) + 
\frac{1}{\tau_2}\left(f^{ {\rm eq  } }_{Si}
-f^{\star   }_{Si}\right).
\label{dis_sol}
 \end{equation}
 We choose  D2Q9 model with nine discrete velocities ${\pmb c}_{i}^{S}$ $(i=0,\cdots,8)$  given as
 \begin{equation}
{\pmb c}_i^S = c^S\begin{cases} (0,0) &\mbox{if } i = 0 \\
 \left(\cos{\frac{(i-1)\pi}{4}},\sin{\frac{(i-1)\pi}{4}}\right)& \mbox{if } i=1,2,3,4\\
 \sqrt{2}\left(\cos{\frac{(i-1)\pi}{4}},\sin{\frac{(i-1)\pi}{4}}\right) &\mbox{if } i =5,6,7,8,
 \end{cases}  
\end{equation}
with the following weights 
\begin{equation}
w_i = \begin{cases} \frac{4}{9} &\mbox{for } i = 0 \\
  \frac{1}{9}& \mbox{for } i=1,2,3,4\\
 \frac{1}{36} &\mbox{for } i =5,6,7,8.
 \end{cases}  
\end{equation}
The lattice sound speed $c_{sS}$ is related to the magnitude of discrete velocity $c^S$ as $(c^S)^2=3 c_{\rm sS}^2$.
 The macroscopic observables, such as mass density, $\rho_S$, momentum density ${\pmb J}_{S}(\rho_s {\pmb u}_S)$ , and stress tensors
 ${\pmb P}_S$ are defined as $\sum_{i}f_{Si}\{1, {\pmb c}_{i}^{S},  {\pmb c}_{i}^{S}  {\pmb c}_{i}^{S}\}=\{\rho_s,{\pmb J}_{S}, {\pmb P}_S\}$. 
 The discrete form of equilibrium distribution function takes the following form \cite{qian1992lattice,shan1998discretization,higuera1989lattice}
\begin{align}
 %\label{Max_boltz}
\begin{split}
\label{dis_eq}
{f}_{Si}^{\rm eq} &= w_i\rho_S\Biggl[1 +
\frac{ {\pmb c}_{i}^S\cdot{\pmb U}}{ \, c_{sS}^2} 
+ \frac{({\pmb c}_{i}^S\cdot{\pmb U})^2}{2\, \,   c_{sS}^4} 
-\frac{({\pmb U}\cdot{\pmb U})}{2\, \,   c_{sS}^2} 
  \Biggr].
 \end{split}
\end{align}
Here, recall ${\pmb U}$ is the total velocity of the solution.
This is an approximate expression and can be improved if needed. Depending on the collision-model, quasi-equilibrium takes different form\cite{arcidiacono2006simulation,arcidiacono2007simulation}. The one where component  momenta are quasi conserved variables, it take the following form
  \begin{align}
  \begin{split}
\label{dis_eq}
{f}_{Si}^{\star} &= w_i\rho_S\Biggl[1 +
\frac{ {\pmb c}_{i}^S\cdot{\pmb u}_S}{ \, c_{sS}^2} 
+ \frac{({\pmb c}_{i}^S\cdot{\pmb u}_S)^2}{2\, \,   c_{sS}^4} 
-\frac{({\pmb u}_S\cdot{\pmb u}_S)}{2\, \,   c_{sS}^2} 
  \Biggr].
 \end{split}
\end{align}
 whereas the one with component stress tensors as  quasi-conserved variables,is 
   \begin{align}
  \begin{split}
\label{dis_eq}
{f}_{Si}^{\star} &= w_i\Biggl[\rho_S +
\rho_S\frac{ {\pmb c}_{i}^S\cdot{\pmb U}}{ \, c_{sS}^2} 
+({\pmb P}_{S}-\rho_S c_{sS}^2{\pmb \delta}):({\pmb c}_{i}^S{\pmb c}_{i}^S -c_{sS}^2{\pmb \delta})
  \Biggr].
 \end{split}
\end{align}
The continuum quantity of the mixture like total mixture velocity, ${\pmb U}$, is calculated using the information from polymer phase, the 
 discrete model of which is described in the subsequent section.
\subsection{\label{lb_polymer}Lattice Boltzmann model  for polymer}
\begin{figure}
\includegraphics[scale=0.5]{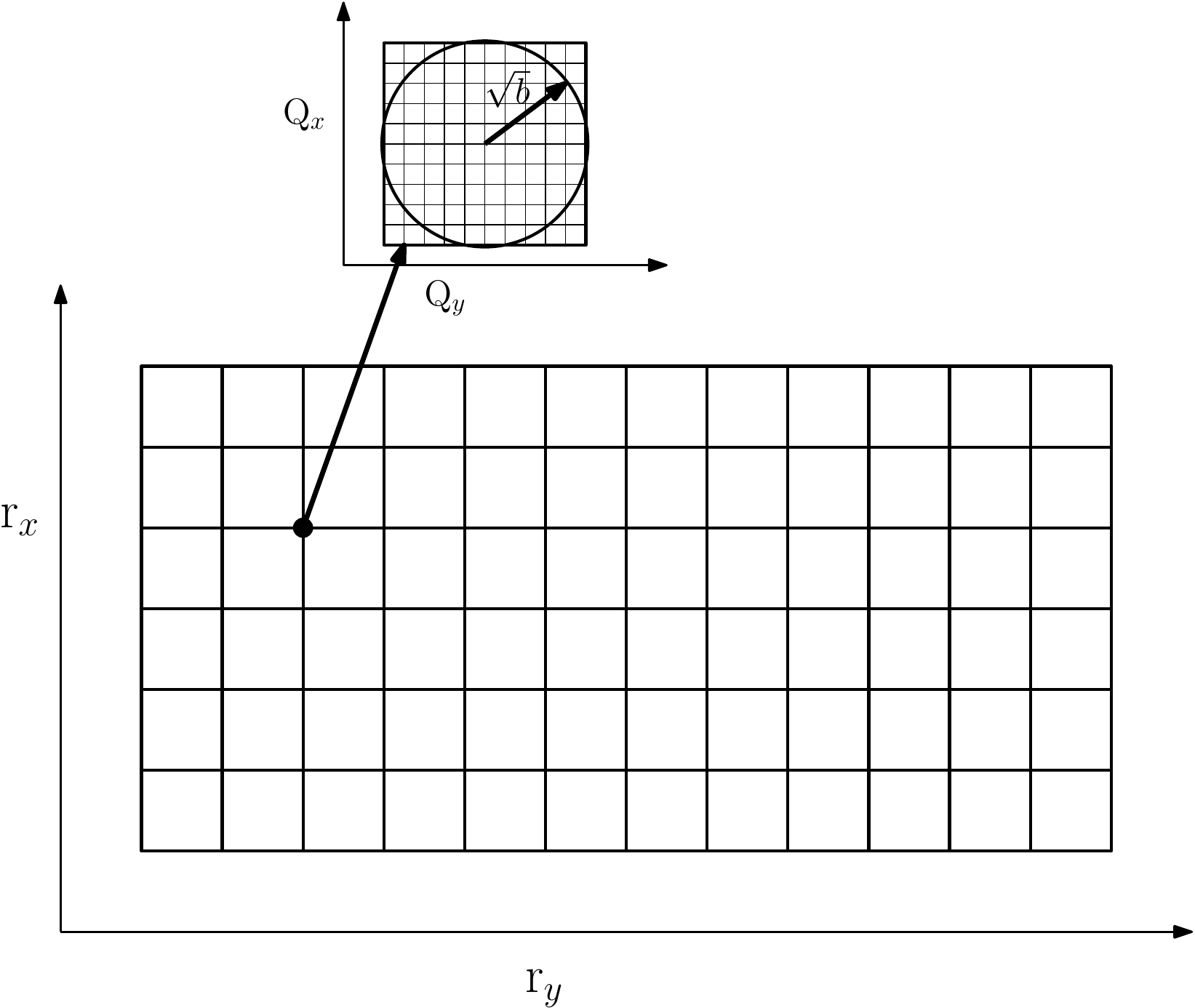}
\caption{Four dimensional configuration space for polymer dumbbell}
\label{4Dconf_space}
\end{figure}

 We first recall the   distribution function for polymer which does not differentiate between the location of the two beads for the numerical convenience, as
 \begin{equation}
 f_P({\pmb r},{\pmb Q},{\pmb v}_{\rm P},\dot{\pmb Q},t)=\sum_{\nu}f^{II}({\pmb r}-{\pmb R}_\nu,{\pmb Q},{\pmb v}_{\rm P}-\dot{\pmb R}_\nu,\dot{\pmb Q},t)
\end{equation}
For the polymeric solute, to solve
a two-dimensional problem in position-orientation space (the orientation being characterized by a single angle), we need to resolve a four-dimensional 
${\pmb r}-{\pmb Q}$ space as shown in Fig. \ref{4Dconf_space}. 
We chose to work with D4Q25  velocity model  
 whose discrete velocities $(c_{i1}^{P},c_{i2}^{P},c_{i3}^{P}, c_{i4}^{P})\equiv(v_{ix},v_{iy},\dot{Q}_{ix}, \dot{Q}_{iy})$ 
are given in Table \ref{dis_vel}.
\begin{table}
\begin{center}
    \begin{tabular}{ | l  | l  |l|l |}
     \hline $r_x$& $r_y$  & $Q_x$   &$Q_y$    \\ \hline
      0 &   0  &0   &0   \\ \hline
      $\pm {\rm c}^P$ &   $\pm {\rm c}^P$  &0   &0   \\ \hline
      $\pm {\rm c}^P$ &  0  & $\pm {\rm c}^P$   &0   \\ \hline
       $\pm {\rm c}^P$ &  0  & 0   &$\pm {\rm c}^P$  \\ \hline      
 0& $\pm {\rm c}^P$  &   $\pm {\rm c}^P$   &0   \\ \hline
 0& $\pm {\rm c}^P$  &  0   & $\pm {\rm c}^P$   \\ \hline
 0& 0 &$\pm {\rm c}^P$   & $\pm {\rm c}^P$   \\ \hline
 %0&0  &0 & $\pm {\rm c}^P$    \\ \hline   
 %(1,-1,0,0)& (1,0,-1,0)  &(1,0,0,-1)   &(0,1,-1,0)   & (0,1,0,-1)    & (0,0,1,-1)\\ \hline  
%(-1,1,0,0)& (-1,0,1,0)  &(-1,0,0,1)   &(0,-1,1,0)   & (0,-1,0,1)    & (0,0,-1,1)\\ \hline  
%(-1,-1,0,0)& (-1,0,-1,0)  &(-1,0,0,-1)   &(0,-1,-1,0)   & (0,-1,0,-1)    & (0,0,-1,-1)\\ \hline  
\end{tabular}
\caption{Discrete velocity set} 
\label{dis_vel}
  \end{center}
\end{table}
% are $(0,0,0,0),(\pm c,\pm c,0,0), (\pm c,0,\pm c,0), (\pm c,0,0,\pm c) 
% (0,\pm c\pm c,0), (0,\pm c,0\pm c)$ and $(0,0\pm c,\pm c)$
Using the following conditions, 
\begin{align}
 \begin{split}
  \sum_{i} w_i=1,\quad   \sum_{i} w_i c_{i\alpha}^{P}c_{i\beta}^{P}=  c_{\rm sP}^2\delta_{\alpha\beta},\quad
  \sum_{i} w_i c_{i\alpha}^{P}c_{i\beta}^{P}c_{i\gamma}^{P}c_{i\theta}^{P}=  c_{\rm sP}^2(\delta_{\alpha\beta}\delta_{\gamma\theta}+
  \delta_{\alpha\gamma}\delta_{\beta\theta}+\delta_{\alpha\theta}\delta_{\gamma\beta}),
%\quad \sum_{i} w_i c_{i\alpha}c_{i\beta}c_{i\gamma}c_{i\theta}=c_{\rm sP}^4\Delta_{\alpha\beta\gamma\theta},
 \end{split}
\end{align}
 the associated weights can be  found as  $w_0=1/3$  and $w_{1-24}=1/36$ with $(c^P)^2=3 c_{\rm sP}^2$ where $c_{\rm sP}$ is the lattice sound speed for D4Q25 model. 
%  {\bf \color{red} GS: This is not very clear. I think the information that you want to convey is that the set of vectors only need to be isotropic until second order, and no more, and set chosen achieves this in four dimensions.
% 
% But, the message doesn't get across....}
The evolution of discrete population is given as
 \begin{align}
 \begin{split}
 \left(\frac{\partial }{\partial t}
+ { v_{i\alpha}}\frac{\partial }{\partial { r_{\alpha}}}+
{ \dot{{Q}}_{i\alpha}}\frac{\partial }{\partial { Q_{\alpha}}}\right)&f^{\rm   }_{Pi}({\pmb r},{\pmb Q},{\pmb v}_{\rm P},\dot{\pmb Q},t)
=
\frac{1}{\tau_1}\left(f^{ \star   }_{Pi}  
-f^{\rm   }_{Pi}\right) + 
\frac{1}{\tau_2}\left(f^{ {\rm eq} }_{Pi} 
-f^{\star \rm   }_{Pi}\right),
\label{dis_LB_r_q_space}
\end{split}
\end{align}
% where a superscript $\nu$ of a function $\phi$ denotes   either of bead location i. e. $\phi^\nu=\phi({\pmb r}-{\pmb R}_\nu,{\pmb Q},{\pmb v}_{\rm P}-\dot{\pmb R}_\nu,\dot{\pmb Q},t)$.
%where $\psi=\sum_if_i$ and  ${\pmb v}_r,{\pmb v}_Q $ are given by Eq. \eqref{impos_inhomo}.
The moments in conformation (${\pmb r}-{\pmb Q}$) space  are defined as 
$\sum_{i}f_{Pi}^{\rm  }\{1, {\pmb v}_{i}, \dot{\pmb Q},{\pmb v}_{i}{\pmb v}_{i}, {\pmb v}_{i}\dot{\pmb Q},\dot{\pmb Q}\dot{\pmb Q}\}=\{\psi,{\pmb J}^{r},{\pmb J}^{Q}, {\pmb P}^r, {\pmb P}^{rQ},  {\pmb P}^Q\}$.
% The first two moments are given as: 
% \begin{align}
%  \begin{split}
%   \psi&=\sum_i{f^{  II }_{Pi}} \qquad
%    { \pmb J}^{r}=\sum_i{f^{  II }_{Pi}} {\pmb v}_{i}, \qquad
%     {\pmb  J}^{Q}=\sum_i{f^{  II }_{Pi}} \dot{\pmb Q}_{i}, 
%   \end{split}
% \end{align}
The  discrete equilibrium distribution can be expressed to linear order as 
\begin{align}
\begin{split}
\label{feq_rq_space}
f^{ {\rm eq}   }_{Pi}&= w_i\Biggl[\psi +
\frac{  {\pmb J}^{r}_{\rm eq}\cdot  {\pmb v}_{ i}}{ c_{sP}^2} +
\frac{  {\pmb J}^{Q}_{\rm eq}\cdot  \dot{\pmb Q}_{ i}}{  c_{sP}^2}   
%+ \frac{\bar{ v}^{\star}_{\alpha} \bar{ v}^{\star}_{\beta}}{2\, c_s^4}  \left(   \dot{{Q}}_{i\alpha}\dot{{Q}}_{i\beta} -  c_s^2 \delta_{\alpha \beta}   \right)  
  \Biggr],
\end{split}
\end{align}
where the value of ${\pmb J}^{r}_{\rm eq}$ and ${\pmb J}^{Q}_{\rm eq}$ is given by Eq. \eqref{poly_imposed_vel}.
The  quasi-equilibrium distributions will take the following form 
\begin{align}
\begin{split}
\label{fstar_rq_space}
f^{\star   }_{Pi}&= w_i\Biggl[\psi +
\frac{  {\pmb J}^{r}\cdot  {\pmb v}_{ i}}{  c_{sP}^2} +
\frac{  {\pmb J}^{Q}\cdot  \dot{\pmb Q}_{ i}}{ c_{sP}^2}   
%+ \frac{\bar{ v}^{\star}_{\alpha} \bar{ v}^{\star}_{\beta}}{2\, c_s^4}  \left(   \dot{{Q}}_{i\alpha}\dot{{Q}}_{i\beta} -  c_s^2 \delta_{\alpha \beta}   \right)  
  \Biggr],
\end{split}
\end{align}
for the collision model with competent momenta as quasi conserved quantify whereas it can be expressed as 
 \begin{align}
  \begin{split}
\label{dis_eq}
f^{\star  }_{Pi} &= w_i\Biggl[\psi+
\frac{  {\pmb J}^{r}_{\rm eq}\cdot  {\pmb v}_{ i}}{  c_{sP}^2} +
\frac{  {\pmb J}^{Q}_{\rm eq}\cdot  \dot{\pmb Q}_{ i}}{ c_{sP}^2} 
+({\pmb P}^{r}-\psi c_{sP}^2{\pmb \delta}):({\pmb v}_{i}{\pmb v}_{i} -c_{sP}^2{\pmb \delta})
+{\pmb P}^{rQ}:({\pmb v}_{i}\dot{\pmb Q}_{i})
+({\pmb P}^{Q}-\psi c_{sP}^2{\pmb \delta}):(\dot{\pmb Q}_{i}\dot{\pmb Q}_{i} -c_{sP}^2{\pmb \delta})
  \Biggr].
 \end{split}
\end{align}
for the one with component stress tensor as quasi conserved variable.
% Again, the quasi-equilibrium distributions will depend on the models. For the model where component momentun is quasi conserved variable, it has the form
% 
% and for the one which has quasi conserved variable as  it has following form.

\subsection{\label{discrete_for_4D_1} Time discretization}
This section reviews   the time discretzation scheme. 
In the lattice Boltzman scheme, Eqs. \eqref{dis_sol},\eqref{dis_LB_r_q_space} are discretized in time by applying the implicit trapezoidal rule between time $t$ as
% $t+\Delta t$ as
% On discretizing Eq. with respect to time, using the trapezoidal rule for evaluating the collision term, 
% we get the evolution equation in implicit form as
\begin{equation}
f_{ji}({\bf x}+{\bf c}\Delta t,t+\Delta t)=f_{ji}({\bf x},t)+\frac{\Delta t}{2}\left[\Omega_j(f_{ji}({\bf x},t))+\Omega_j(f_{ji}({\bf x}+{\bf c}^j\Delta t,t+\Delta t))\right]\label{eq1}
\end{equation}
where, $j=S,P$ and $\Omega_{S,P}$ represents the collision operator for solvent/polymer \cite{chen_annual_rev}.
In order to make the method explicit, following auxiliary function, $g_{ji}$, is introduced which depends on original distribution function, $f_{ji}$, as 
\begin{equation}
\label{aux}
 g_{ji} =f_{ji}-\frac{\Delta t}{2}\left[\frac{1}{\tau_1}\left(f_{ji}-f_{ji}^{\rm \star}\right)+\frac{1}{\tau_2}\left(f_{ji}^{\rm  \star}-f_{ji}^{\rm \rm eq}\right) \right], 
\end{equation}
After the transformation, the  resultant discrete equation becomes
\begin{equation}
g_{ji}(  {\bf x}+{\pmb c}^{j}\Delta t,t+\Delta t)=g_{ji}(  {\bf x},t)\left( 1-  2\beta \right) +2\beta\left[\left( 1-  \frac{\tau_1  }{\tau_2}\right)f_{ji}^{\rm \star}   
+   \frac{\tau  }{\tau_1}   f_{ji}^{\rm eq }  \right],
\end{equation}
 where $\beta=\Delta t/(2\tau_1+\Delta t)$. Since, $g_{ji}$ depends on the both $f^{\star}_{ji}$ and $f^{\rm eq}_{ji}$, the collision model require the evaluation of the moments of in term of $ f_{ji}$. Therefore,
\begin{align}
 \begin{split}
\rho_S(f_S)=\rho_S(g_S), &\qquad \psi(f_P^{\rm})=\psi(g_P);\\
{\pmb J}_S(f_S)= \frac{\frac{2\tau_{1,2}}{\Delta t}{\pmb J}_s(g_S)+{\rho_s}{\pmb U}}{1+\frac{2\tau_{1,2}}{\Delta t}}, & \qquad
{\pmb J}_r(f_P^{\rm })= \frac{\frac{2\tau_{1,2}}{\Delta t}{\pmb J}_r(g_P)+{\pmb J}^{\rm r}_{\rm eq} }{1+\frac{2 \tau_{1,2}}{\Delta t}},
 \end{split}
\end{align}
It is worth mentioning at this point that in order to calculate polymeric contribution to the total velocity, ${\pmb U}$, we need to further integrate out the conformation dependence of polymer momentum density. Therefore, the dependence of the transformation into the auxiliary function, ${\pmb g}$ , on total moments looks like  $\rho(f)=\rho(g) $ and 
\begin{equation}
 {\pmb U}(f)={\pmb U}(g)+(\Delta t/2\tau_{1,2}){\pmb \Theta(g)}/{\rho}.
\end{equation}
Finally, the initial condition on $\psi({\pmb r},{\pmb Q},t)$ at every location in ${\pmb r}$ is given as
\begin{equation}
 \psi({\pmb r}, {\pmb  Q},0)=
\begin{cases} N_{\rm eq}(1-Q^2/b)^{b/2} & \text{for  $|Q| \leq \sqrt{b}$}
\\
0 &\text{ elsewhere,}
\end{cases}
\label{initail_condition}
\end{equation}   
where $N^{\rm eq}=2\pi b^{3/2}B\{3/2, (b + 2)/2\}$  and $B\{x, y\}$ is the Beta function.
The FENE spring force has a singularity at $Q=\sqrt{b}$ for limiting the  maximum extension of the spring upto a length of $\sqrt{b}$. The simulation domain in Q-space is limited inside a circle of radius $\sqrt{b}$ as shown in Fig. \ref{4Dconf_space}.  The bounce-back boundary condition is applied at the boundaries of the circle \cite{chen_annual_rev,succi_book,ladd1994numerical}.

% of 
\section{\label{numerical_validation_kolm}Viscoelastic Kolmogorov Flow}
In this section, we validate the kinetic theory formulation detailed in the earlier sections by showing that the  presented model is capable of capturing the viscoelastic effects exhibited by polymer solutions.
We choose the Kolmogorov flow for this purpose. In this flow, a unidirectional body force varying sinusoidally in space, and represented as $ {\pmb f}=\left[F\cos\left({{y}/{l}}\right),0\right]$, is used  to induce a parallel flow with velocity  $ U\cos(y/l) $. The magnitude of the force is then given as $F=\eta U/l^2$, with $\eta$ being the viscosity.

The Newtonian Kolmogorov flow becomes linearly unstable for Reynolds number (Re) greater than $\sqrt{2}$ \cite{meshalkin1961investigation}, the essentially inviscid instability arising from the presence of inflection points in the base-state sinusoidal velocity profile. For the case of a dilute polymer solution, effects of elasticity have been shown to stabilize the Newtonian inflectional instability associated with a shear layer \cite{azaiez1994linear}.
%[Azaiez & Homsy reference 72].
The stabilization arises because the stretched polymers lead to the perturbed shear layer acting as a deformed elastic membrane, and the resulting restoring force leads to the damping of short-wavelength perturbations. Subsequent efforts \cite{boffetta2005viscoelastic,berti2008two,garg2018viscoelastic,khalid2020center,dubief2020first} have examined the susceptibility of Kolmogorov flow, and other wall-bounded uni-directional shearing flows, to elastoinertial instabilities. Very recently, elasticity alone has been shown to destabilize a uni-directional shearing flow \cite{khalid}, even in the absence of inertia. The mechanism underlying the aforementioned elastoinertial and purely elastic instabilities is currently under examination, and the subsequent nonlinear evolution is therefore beyond the scope of the present numerical investigation.

For purposes of numerically verifying the stabilizing action of elasticity on Kolmogorov flow, we consider a unit cell in two-dimensional physical space, of side $2\pi$, discretized using 72 grid points. We use
32 grid points to discretize the conformation space.
Periodic boundary conditions are used in both spatial directions for the solvent as well as the polymer solver. 
A Gaussian random field is used to seed the instability in the flow. 
In our study, we use $l = 1/4$, implying that the unit cell incorporates four periods of the Kolmogorov profile. 
The Reynolds Number, Re, is defined using the kinematic viscosity of solution, $\nu$, as $Ul/\nu$. The additional physical parameters needed for the viscoelastic case are as follows. The first parameter is $\beta$ which represents the ratio of the solvent viscosity $\eta_s$ to the solution viscosity ($\eta_s+\eta_p$), with the polymeric contribution to the viscosity  $\eta_p=n_{p}k_{\rm B}T\tau_{\rm R}$; here, $n_p$ is the polymer  number density. 
Next, we have the Weissenberg number defined as  $Wi= U\tau_{\rm R}/l$.  Finally, for the FENE dumbbells used to represent the polymer molecules, the maximum   extensibility parameter, $b$ is set to be 25.  With the choice of the other parameters used in the present study, the value of Sc lies between 0.05 and 4.0. 
To explore the elastic effect of the polymer on the flow, we have considered a  scenario where the flow is unstable to infinitesimal amplitude perturbations in the Newtonian limit (Re=3.5).

\begin{figure*}[h]
     \centering
         \includegraphics[width=1.\textwidth, height=10cm]{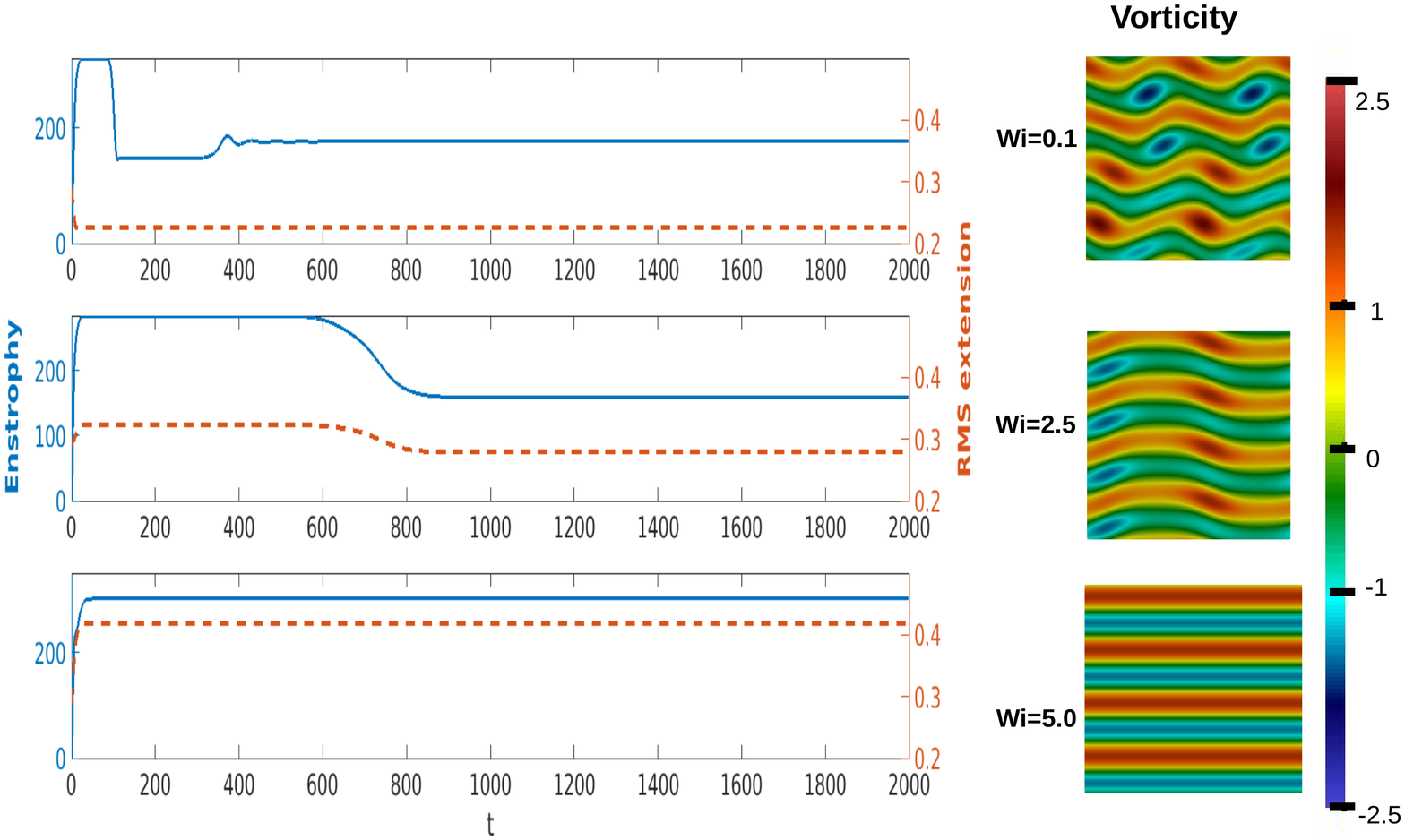}
         \caption{Time dynamics of enstrophy, RMS-extension and corresponding vorticity profile at convective time, t= 2000 at Re=3.5 and $\beta=0.5$ with varying Wi.}
         \label{re3p5}
\end{figure*}
Fig.\ref{re3p5} shows the  vorticity  fields  (${\pmb \omega}=\nabla\times {\pmb u}$ with ${\pmb u}$ being the flow velocity) characterizing the saturated nonlinear state, and the temporal development of global quantities - both the enstrophy (defined as $1/2\int ({\pmb \omega}\cdot{\pmb \omega}) \, d{x} d{ y}$) and the root mean square extension of the polymer (defined as $Q_{\rm rms}({\pmb x},{\pmb y},t)=
 %\sqrt{<{Q^2}({\pmb x},{\pmb y},t)}>=
 \sqrt{{<\psi Q^2>}/{<\psi>}}, $ where $\psi$ is the conformation probability density and can be understood as the zeroth moment (i.e. $\psi=\sum_i f^{P}_i$) of the two particle distribution function characterizing the  polymer molecule. The temporal development of the enstrophy may be explained as follows. On short time scales, momentum diffusion arising from the induced forcing leads to the laminar sinusoidal velocity profile for all three Wi examined. For the two smaller Wi's, there is a decrease in the enstrophy on longer time scales, corresponding to the onset of the inflectional instability mentioned above. The onset of instability, and the associated velocity fluctuations at the chosen Re lead to a higher rate of viscous dissipation, in turn leading to a mean profile that is still nearly sinusoidal but with a smaller amplitude. This smaller amplitude leads to a lower enstrophy, and is responsible for the aforementioned decrease in enstrophy. Note that this decrease happens on a shorter time scale for Wi = 0.1 owing to the instability having a nearly Newtonian character. For Wi = 2.5, the decrease is delayed, and has a marginally smaller magnitude, reflecting an elasticity-induced stabilization. The corresponding vorticity field plot shows that the saturated state for Wi = 2.5 is characterized by a larger length scale in the streamwise direction; this increase in the characteristic length scale is consistent with the tendency of the stresses arising from stretched polymers acting to damp out the shorter wavelength perturbations arising from an inflectional instability \cite{azaiez1994linear}. On increasing Wi to 5, the instability disappears, which is likely due to the dominant unstable modes shifting to wavelengths that are larger than the size of the periodic domain; correspondingly, the enstrophy remains at the plateau value, corresponding to the laminar profile, for all time.
The plots of the root mean square polymer extension field reflect the trends in the enstrophy variation mentioned above. 
 
\begin{figure}[h]
\subfigure[t=70]{\includegraphics[width=0.5\textwidth]{{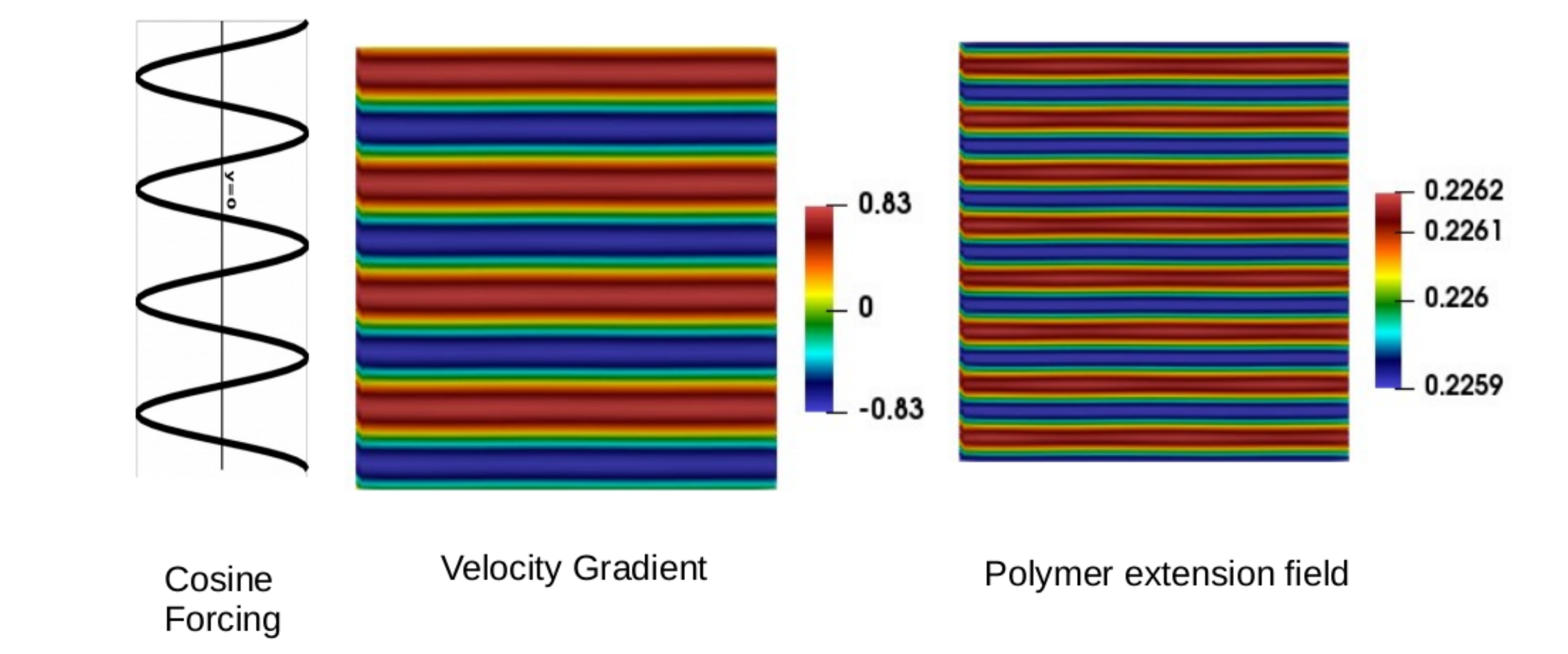}} \label{3p5_wp1_before}}
\subfigure[t=90]{\includegraphics[width=0.45\textwidth]{{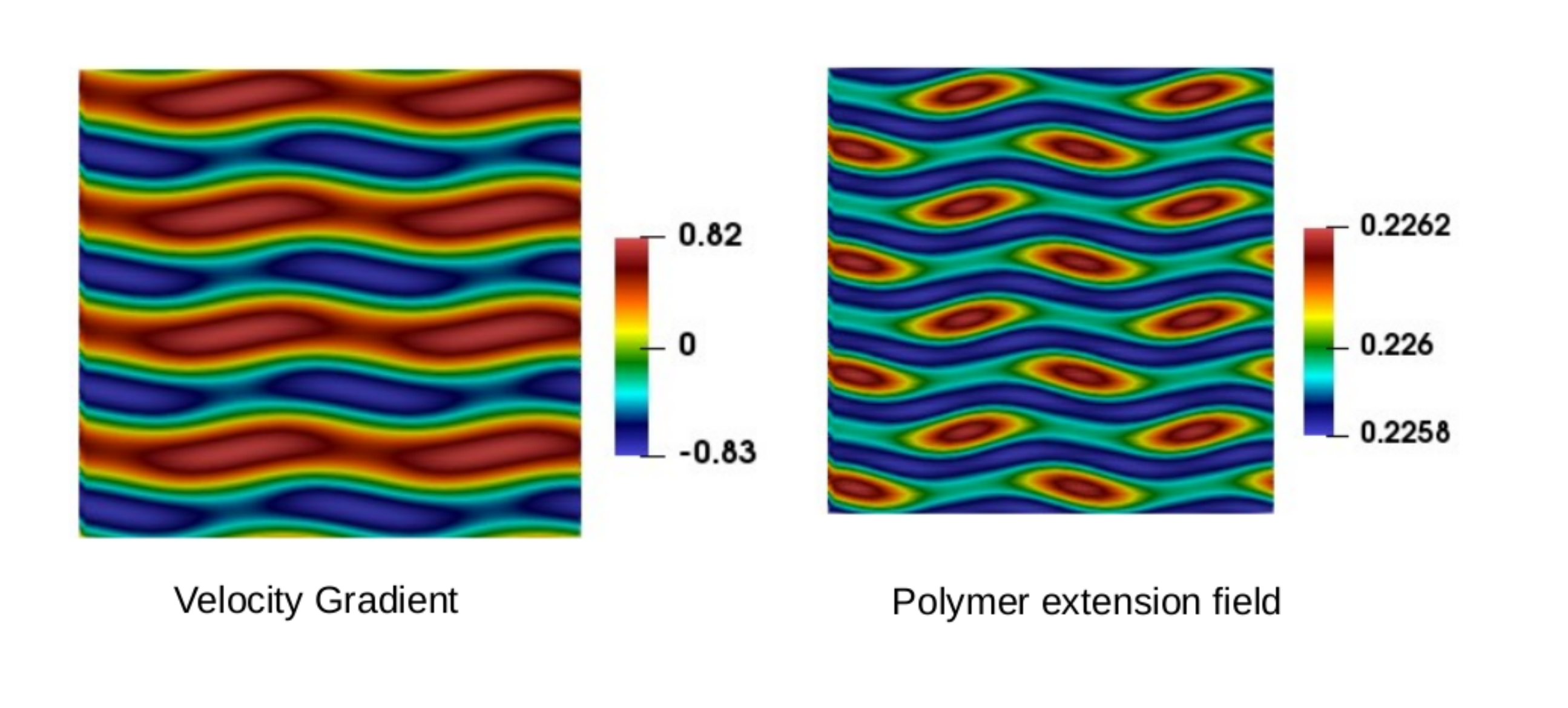}} \label{3p5_wp1_after}}
\subfigure[Saturated state (t=2000)]{\includegraphics[width=00.45\textwidth]{{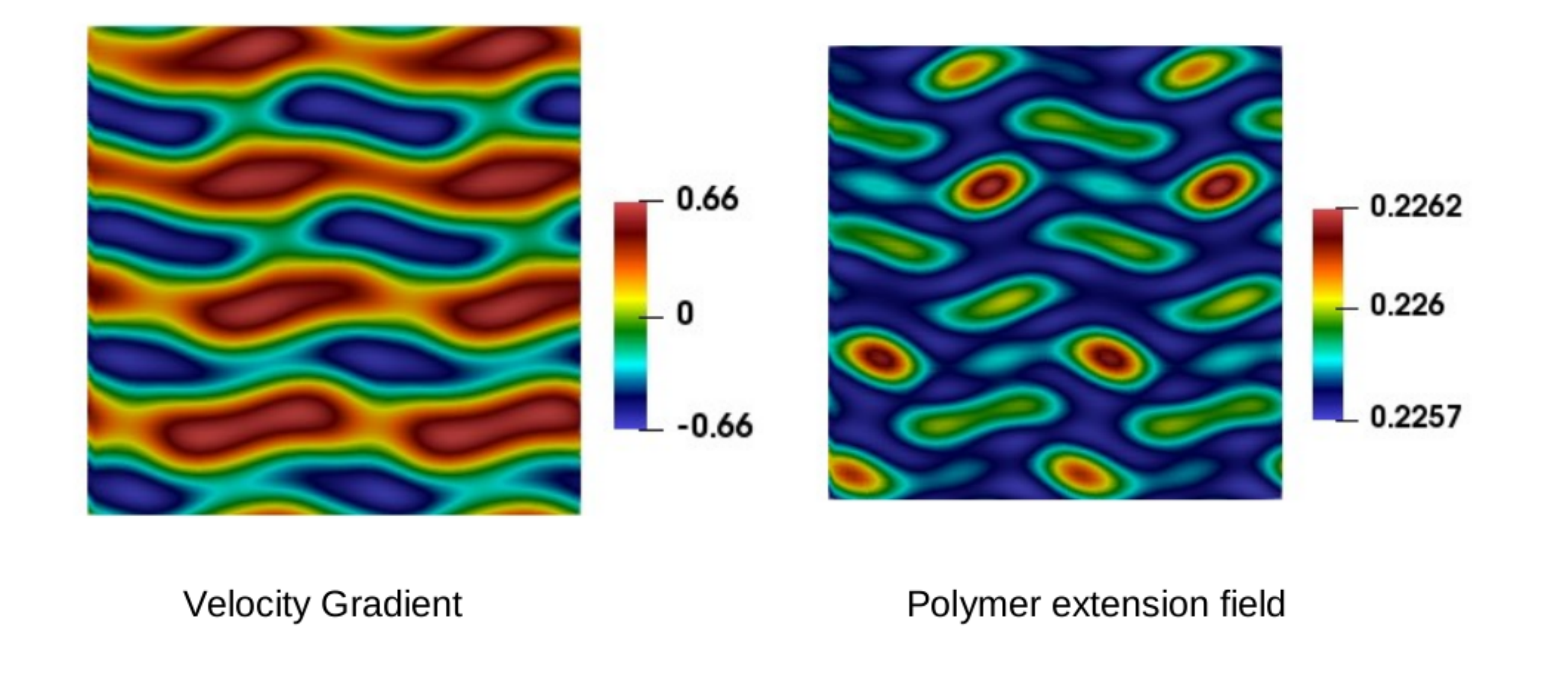}} \label{3p5_wp1_sat}}
\caption{Spatial distribution of global quantities at Re=3.5 and Wi=0.1 at different time.}
\label{Re_3p5_p1f}
\end{figure}

In order to understand what actually happens due to the non-homogeneity of flow on polymer, we first consider the case for Wi=0.1. For this low Wi, the polymer feedback to the flow will be minimal. Figure 9a, b and c show the scenario just before and after the instability, and for much longer times corresponding to the nonlinear saturated state. To orient the reader, in Fig \ref{3p5_wp1_before}, we have also plotted the cosine forcing over four periods, used to initiate the Kolmogorov flow. Before instability onset, at t=70, the figure shows the expected one-dimensional variations of the velocity gradient and polymer extension fields, with the greatest extensions correlating to the maximum values (both positive and negative) of the velocity gradient. At t=97, when the inflectional instability has just developed, one starts to observe the emergence of two-dimensional variations in the aforementioned fields in Fig \ref{3p5_wp1_after}. Finally, the saturated two-dimensional fields are shown in Fig \ref{3p5_wp1_sat}, corresponding to t = 2000. For the small Wi chosen, one notes the modest of ${\rm Q}_{\rm rms}$ which fluctuates around 0.2 which is the equilibrium extension value, $l_0$ for the chosen parameter b (i.e. $1/\sqrt{25}$).

\begin{figure}[h]
\subfigure[t=400]{\includegraphics[width=0.45\textwidth]{{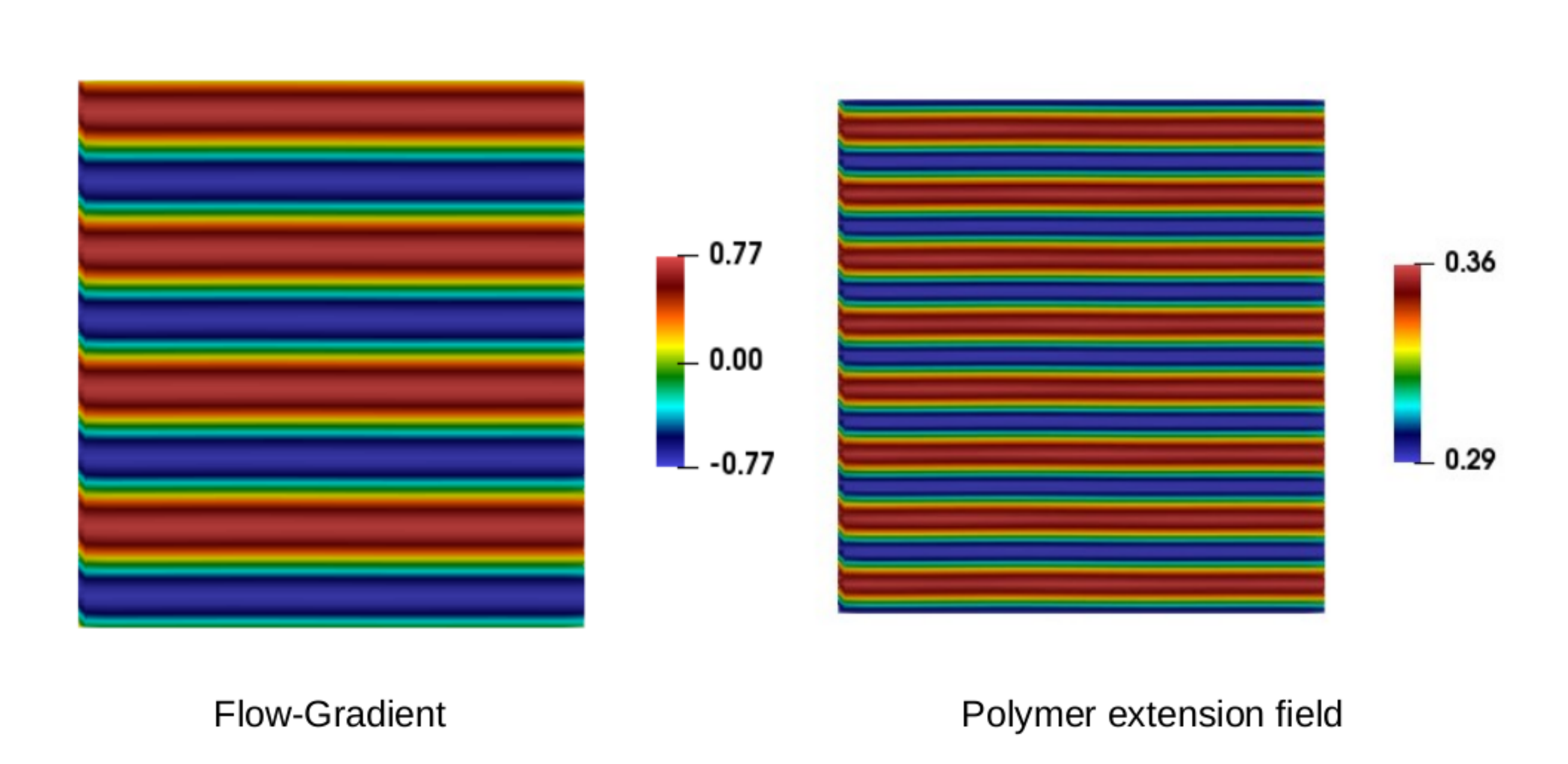}} \label{3p5_w2p5_before}}
\subfigure[t=700]{\includegraphics[width=0.45\textwidth]{{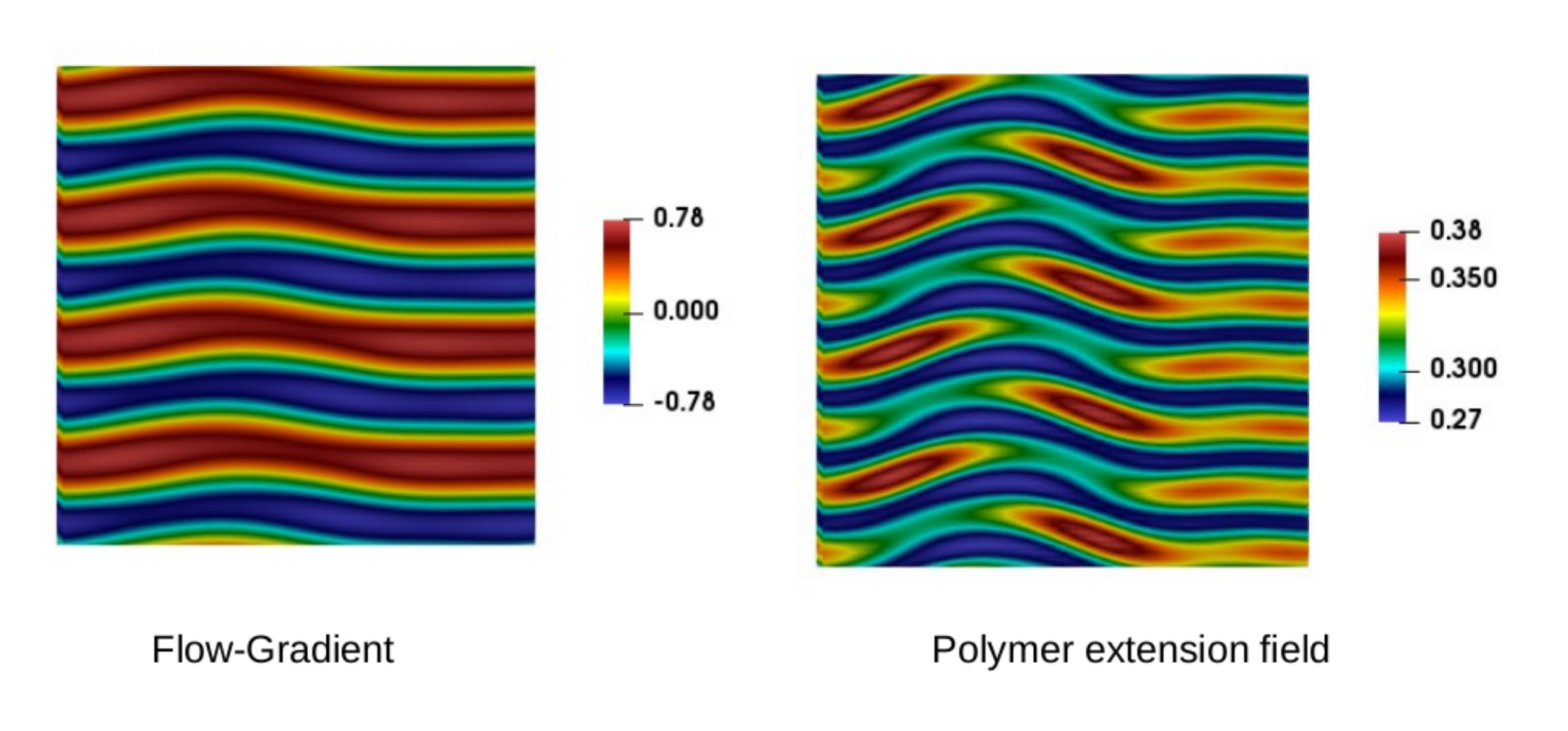}} \label{3p5_w2p5_after}}
\subfigure[Saturated state (t=2000)]{\includegraphics[width=00.45\textwidth]{{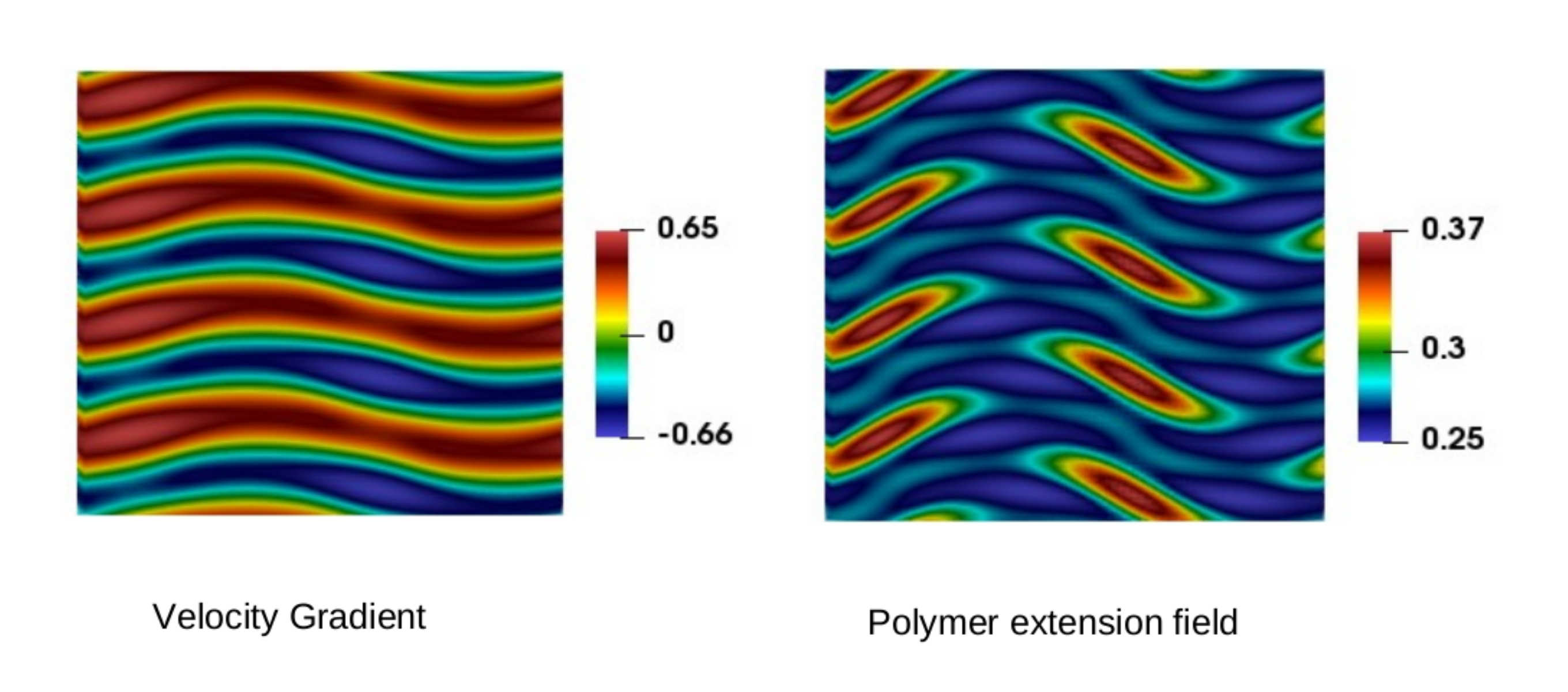}} \label{3p5_w2p5_sat}}
\caption{Spatial distribution of global quantities at Re=3.5 and Wi=2.5 at different time.}
\label{Re_3p5_2p5}
\end{figure}

Similar to the previous case, for  Wi=2.5, we again considered three time instances- just before and after the instability and at a long time corresponding to the nonlinear saturated state in Fig. \ref{Re_3p5_2p5}. At t=400, Fig. \ref{3p5_w2p5_before} shows that  the flow remain in the base-state with one dimensional sinusoidal variation where polymer extension  aligns with the flow. We also see that the polymer extension is fluctuating between the values   higher than that of  Wi=0.1 case  which resulted in higher $Q_{\rm rms}$ (see Fig.\ref{re3p5}). At t=700, the velocity gradient  shows onset of instability indicated  by the deviation  from the laminar shape.   The polymer reorients accordingly and the value of $Q_{\rm rms}$ starts to sharpen in the region of extreme value of velocity gradient in Fig. \ref{3p5_wp1_after}. Figure \ref{3p5_wp1_sat} shows that the saturated instability in the velocity gradient, as mentioned earlier,  is now characterized by a larger length scale in the streamwise direction and the  polymer extension is now fully concentrated in the extremum of flow gradient 
%and the RMS extension value reaches another steady state (Fig. \ref{re3p5}).
 At Wi=5.0, the presence of polymer helps in complete reduction of inertial instabilities and the flow remains stable which is possibly due to the dominant unstable modes being shifting to wavelengths that are larger than the domain size. The velocity gradient profile and polymer extension distribution are similar to the case of stable laminar flow which are described in detail using Fig. \ref{3p5_wp1_before}, however,  $Q_{\rm rms}$ attains a value close to 0.5 (see Fig. \ref{re3p5}) which indicates that the mean extension reaching nearly twice the equilibrium extension is sufficient enough to completely dampen the inertial instabilities.

\section{Conclusion\label{conclusion4d}}
We present a kinetic level coupling of polymer-solvent system in velocity phase space using a Boltzmann-type collision dynamics of mixture to uncover the polymer dynamics. Unlike binary gas mixtures of simple molecules, one encounters an asymmetry in the description of the individual components. This happens because the
polymer dumbbell is represented by a two particle distribution function whereas the solvent phase molecule
is represented by a single particle distribution function. The collision between the polymer dumbbell and
solvent molecule is modeled using a quasi-equilibrium based relaxation collision kernel.
 The detailed kinetic scheme also results in a  continuum picture where dissipative coupling between the phases occurs naturally. The present kinetic formulation also leads to the Smoluchoswki equation which governs configuration  space dynamics. Using this kinetic description, a  numerical algorithm is then built along the lines of lattice Boltzmann method. Finally, via numerical simulation of two dimensional viscoelastic Kolmogorov flow, we are able to recover the canoninal effect of polymer elasticity, particularity, the suppression of inflectional instabilities due to inertia, in turn leading to a saturated nonlinear state characterized by a length scale that increases with increasing Wi.
 %as well as development of elastic instabilities in otherwise laminar flow. 
 In present scheme, the   polymer-solvent coupling occurs in velocity phase space, hence eliminating the need of
any closure approximations. Therefore, this scheme has the potential to advance  our understanding of viscoelastic flow phenomena, including instabilities, particularly in cases where the polymer molecules are represented by realistic micromechanical models, going beyond the Hookean dumbbell/bead-spring representations, that automatically preclude the derivation of closed-form constitutive equations for the polymer stress.

\section{Acknowledgement}
S.S acknowledge the financial support by   EPSRC (UK) grant (EP/N016602/1) and the  Leverhulme Early Career Fellowship.
S.A. thank SERB funding for project ``Multiscale Modeling of complex fluid".
% This work has been financially supported by EPSRC (UK) grant (EP/N016602/1) and the Leverhulme Trust and SERB "Multiscale Modeling of complex fluid".
S.S. acknowledges the use of the Scientific Computing Research Technology Platform, and associated support services at the University of Warwick, in the completion of this work. 
 SS would like to thank James Sprittles and Laura Cooper (University of Warwick,Coventry,  UK), for  helpful  discussions
related to this work.
%\section{\label{appen}Appendix A: Chapman-Enskog Expansion}
\section{\label{appen}Appendix A: Chapman-Enskog Expansion}
In this section, using a multi-scale Chapman-Enskog expansion, it is shown that in  the present BGK type 
collision model (Eq. \eqref{fp_x1_x2}), 
the correct slow dynamics of configuration distribution function is recovered in the dilute limit
for  both homogeneous as well as inhomogeneous case.
 In  the    Chapman-Enskog multi-scale expansion, $f^{\rm II}_{\rm P}$ is expanded as
\begin{equation}
 f_{\rm P}^{\rm II}=f^{\rm eq II} +\tau f_{\rm P}^{(1) II}+
\tau^2 f_{\rm P}^{(2)II}+..... \quad \textrm{such that} \int   f^{(n)II} d\dot{\bf Q}\,d{\pmb v}=0 \quad\textrm{for}\quad n>1,
\end{equation}
 The consequence of this is that
the non-conserved moments  can also be expanded in   powers of smallest time scale $\tau$ ($\tau_{1,2}$ depending on collision model) around their equilibrium values.
For example, the momentum and the second-order moments, ${\pmb M}({ \pmb J}^{\rm r}, { \pmb J}^{\rm Q}, { \pmb P}^{\rm r},{ \pmb P}^{\rm rQ}, { \pmb P}^{\rm Q})$    have the following expansions,
\begin{align}
 \begin{split}
{ \pmb M} &= { \pmb M}^{\rm r eq}  +\tau  { \pmb M} +....\\
\end{split}
\end{align}
% %  which are the momentum and second order moment
% % in the present case.
% \begin{align}
%  \begin{split}
% { \pmb J}^{\rm r} &= { \pmb J}^{\rm r eq}  +\tau  { \pmb J}^{\rm r(1)} +....\\
% { \pmb J}^{\rm Q} &= { \pmb J}^{\rm Q eq}  +\tau  { \pmb J}^{\rm Q(1)} +....\\
% { \pmb P}^{\rm r} &={ \pmb P}^{\rm r eq} +\tau{ \pmb P}^{\rm r (1)} +.......\\
% { \pmb P}^{\rm rQ} &={ \pmb P}^{\rm rQ eq} +\tau{\pmb  P}^{\rm rQ (1)} +.......\\
% { \pmb P}^{\rm Q} &={ \pmb P}^{\rm Q eq} +\tau{ \pmb P}^{\rm Q (1)} +.......\\
%   \end{split}
% \end{align}
where the leading order contribution to  equilibrium values are:
\begin{align}
 \begin{split}
 { \pmb J}^{\rm r eq}  &= \psi({\pmb r} ,{\pmb Q},t){ \pmb U} +\sum_\nu\left(\frac{{  \pmb F}_{\nu}}{\zeta}\psi({\pmb r}-{\pmb R}_{\nu}, {\pmb Q}, t) \right),  \qquad
 {  \pmb J}^{\rm Q eq}  =  \psi({\pmb r} ,{\pmb Q},t){\pmb Q}\cdot\frac{\partial {\pmb U}}{\partial {\pmb r}}-\psi({\pmb r} ,{\pmb Q},t)\frac{2{\pmb F}}{\zeta},\\
 {  \pmb P}^{\rm r eq} &=\psi({\pmb r},{\pmb Q},t)\frac{k_{\rm B}T}{m_{\rm B}}{\pmb \delta},\qquad
%+ \underbrace{\psi({\pmb r},{\pmb Q},t){  U}_\alpha{  U}_\beta
%+\sum_\nu\left[\psi({\pmb r}-{\pmb R}_\nu,{\pmb Q},t)\left(U_\alpha\frac{F_{\beta\nu}}{\zeta}+U_\beta\frac{F_{\alpha\nu}}{\zeta}
%+\frac{F_{\alpha\nu}}{\zeta}\frac{F_{\beta\nu}}{\zeta}\right)\right]}_{D_{\alpha\beta}({\pmb r},{\pmb Q},t)}\\
 { \pmb  P}^{\rm rQ eq} =\sum_{\nu}(-1)^\nu\psi({\pmb r}-{\pmb R}_\nu,{\pmb Q},t){\pmb \delta},\qquad
%-\underbrace{\sum_\nu \psi({\pmb r}-{\pmb R}_\nu,{\pmb Q},t)\left({  U}_\alpha+\frac{{  F}_{\alpha\nu}}{\zeta}\right)
%\left((-1)^{\nu}[{  U}_\beta({\pmb r}-2{\pmb R}_\nu,t)-{  U}_\beta({\pmb r},t)]-\frac{2{  F}_{\beta\nu}}{\zeta}\right)
%}_{{  B}_{\alpha\beta}({\pmb r},{\pmb Q},t)}\\
 { \pmb P}^{\rm Q eq}  =2\psi({\pmb r},{\pmb Q},t)\frac{k_{\rm B}T}{m_{\rm B}}{\pmb \delta}.
  \end{split}
\end{align}
The time derivative is also expanded as:
\begin{equation}
 \frac{\partial \phi}{\partial t}= \frac{\partial^{(0)} \phi}{\partial t}+\tau \frac{\partial^{(1)} \phi}{\partial t}
+...
\end{equation}
The moment equation \eqref{polymer_moment_sum} at  the zeroth order  is
\begin{align}
 \begin{split}
 -  { \pmb  J}^{\rm r(1)}&=  \left[
\frac{\partial  }{\partial { \pmb  r}}\cdot \left(\psi({\pmb r},{\pmb Q},t)\frac{k_{\rm B}T}{m_{\rm B}}{\pmb   \delta}\right)
+   \frac{\partial  }{\partial{ {  \pmb  Q}}}{\pmb P}^{\rm rQ\, eq}\right]+ \frac{\partial^{(0)}  }{\partial t} {\pmb J}^{\rm r \,eq},
\\
-   { \pmb  J}^{\rm Q(1)}&=
\left[
\frac{\partial  }{\partial{ {  \pmb r}}}\cdot {\pmb P}^{\rm rQ\, eq}+
\frac{\partial  }{\partial{ {\pmb   Q}}} \cdot \left(2\psi({\pmb r},{\pmb Q},t)\frac{k_{\rm B}T}{m_{\rm B}}{ \pmb \delta}\right) \right] 
+ \frac{\partial^{(0)}  }{\partial t}{\pmb J}^{\rm Q\, eq}.
\end{split}
\end{align}
which gives the configuration distribution evolution as
\begin{align}
 \begin{split}
 \frac{\partial }{\partial t}  {\psi}({\pmb r}, {\pmb Q}, t)  + &  
\frac{\partial  }{\partial{\pmb   r } }  \left(
\psi({\pmb r} ,{\pmb Q},t){ \pmb U} +\sum_\nu\left(\frac{{  \pmb F}_{\nu}}{\zeta}\psi({\pmb r}-{\pmb R}_{\nu}, {\pmb Q}, t) \right) \right)
+ \frac{\partial  }{\partial{  \pmb  Q}}  \left(
 \psi({\pmb r} ,{\pmb Q},t){\pmb Q}\cdot\frac{\partial {\pmb U}}{\partial {\pmb r}}-\psi({\pmb r} ,{\pmb Q},t)\frac{2{\pmb F}}{\zeta}\right)\\
 &=\frac{k_{\rm B}T}{\zeta}\left(\frac{\partial^2\psi}{\partial {\pmb r}^2}+2\frac{\partial^2\psi}{\partial {\pmb Q}^2}\right).
\end{split}
\label{Smoluchowski_new}
\end{align}
  where  $\tau$ is characteristic timescale for velocity fluctuations defined as $\tau=m_{\rm B}/\zeta$ \cite{schieber1988effects,ottinger1996kinetic}.
\subsubsection{Homogeneous  flow in dilute limit}
In dilute limit ${\pmb U({\pmb r},t)\approx {\pmb u}_{\rm S}({\pmb r},t) }$
and for homogeneous flows  the elements
of  velocity gradient tensor ${\pmb \nabla}{\pmb u}_{\rm S}$ can be taken as constant.
Therefore on integrating the ${\pmb r}$ degrees of freedom from Eq. \eqref{Smoluchowski_new}, one gets
%for the case of homogeneous flows:
\begin{align}
 \begin{split}
 \frac{\partial }{\partial t}  {\psi}( {\pmb Q}, t)  + &  \frac{\partial  }{\partial{  \pmb  Q}} \cdot \left(\psi( {\pmb Q},t)\,{\pmb Q}\cdot\frac{\partial {\pmb u}_{\rm S}}{\partial {\pmb r}}
  -\psi({\pmb r} ,{\pmb Q},t)\frac{2{\pmb F}}{\zeta} 
+\frac{2 k_{\rm B} T}{\zeta}\frac{\partial { \psi}}{\partial {\pmb Q}}\right )
=0, 
\end{split}
\label{Smoluchowski_new1}
\end{align}
which is the desired Smoluchowski Equation in the homogeneous flow scenario.
%with $\tau=m_{\rm B}/\zeta$.
 \subsubsection{Density diffusion equation in dilute limit}
 In order to obtain the polymer density equation,
 ${\pmb Q}$ degrees are  integrated out from the  Eq. \eqref{Smoluchowski_new}, which gives
 \begin{align}
 \begin{split}
 \frac{\partial }{\partial t}  {\rho_{\rm P}}({\pmb r},  t)  + &  
\frac{\partial  }{\partial{\pmb   r } }\cdot  [
 \rho_{\rm P}\,{\pmb U}+\frac{m_{\rm B}}{\zeta}\frac{\partial  }{\partial{ \pmb   r }}{ \Theta}_{\alpha\beta} +\frac{m_{\rm B}^2}{\zeta}\int d{\pmb Q } {  J}^{\rm r(1)}_{\alpha}]
=0, 
\end{split}
\end{align}
after multiplying with $m_{\rm B}$. The last term of the above equation is given as 
% Here,
% \begin{align}
%  \begin{split}
%  \hat{C}_{\alpha}({\pmb r} , t)  &=\int d{\pmb Q}  {C}_{\alpha}({\pmb r} ,{\pmb Q} , t)\\
% &=\int d{\pmb Q}\left(\psi({\pmb r},{\pmb Q},t)U_\alpha+\frac{F_\alpha}{\zeta}Q_\beta\frac{\partial}{\partial r_\beta}\psi\right)\\
% &=\rho_{\rm P}\,U_\alpha+\frac{m_{\rm B}}{\zeta}\frac{\partial  }{\partial{    r }_\beta}{ \Theta}_{\alpha\beta}.
% \end{split}
% \end{align}
\begin{align}
 \begin{split}
- m_{\rm B}\int d{\pmb Q } {  J}^{\rm r(1)}_{\alpha}=\left[
\frac{\partial  }{\partial{ {   r}_\beta}} \left(\rho_{\rm P}({\pmb r},t)\frac{k_{\rm B}T}{m_{\rm B}}{  \delta}_{\alpha\beta}\right)
\right]
+ {\frac{\partial^{(0)}  }{\partial t}\left(\rho_{\rm P}({\pmb r},t)U_\alpha\right)} 
\end{split}
\end{align}
%Now, we need to find, $\Xi_\alpha$.
Using the total momentum conservation at macroscopic level 
 \begin{align}
 \begin{split}
 \frac{\partial {\pmb J}}{\partial t}     + 
%\frac{\partial {p}_{\rm P}}{\partial  {\pmb r}} 
\frac{\partial  }{\partial{ {\pmb  r}}}\cdot{{\pmb P}}({\pmb r}, t) 
 &= \frac{\partial }{\partial {\pmb r}}\cdot{\pmb  \Theta}
,\\
\label{poly_sol_mom_app}
\end{split}
\end{align}
we get
% \begin{align}
%  \begin{split}
%  \rho\frac{\partial^{(0)} { U}_\alpha}{\partial t}     + \underbrace{ { U}_\alpha\frac{\partial^{(0)} \rho }{\partial t}}_{\rm higher-order}   
% %\frac{\partial {p}_{\rm P}}{\partial  {\pmb r}} 
%   &=\frac{\partial }{\partial {  r}_\beta}\left( {   \Theta}_{\alpha\beta}-P^{\rm eq}_{\alpha\beta}\right)
% ,\\
%  \frac{\partial^{(0)} { U}_\alpha}{\partial t} &=
% \frac{1}{\rho}\frac{\partial }{\partial {  r}_\beta}\left( {   \Theta}_{\alpha\beta}-n k_{\rm B} T\delta_{\alpha\beta}\right)
% ,\\
% %  \frac{\partial^{(0)} { U}_\alpha}{\partial t} &=
% % \frac{1}{\rho}\frac{\partial }{\partial {  r}_\beta}\left( {   \Theta}_{\alpha\beta}-n k_{\rm B} T\delta_{\alpha\beta} \right).
% \label{poly_sol_mom_appen2}
% \end{split}
% \end{align}
%Therefore,
\begin{align}
 \begin{split}
-\frac{m_{\rm B}^2}{\zeta}\int d{\pmb Q } {  J}^{\rm r(1)}_{\alpha}
% \frac{m_{\rm B}}{\zeta}\left[
% \frac{\partial  }{\partial{ {   r}_\beta}} \left(\rho_{\rm P}({\pmb r},t)\frac{k_{\rm B}T}{m_{\rm B}}{  \delta}_{\alpha\beta} \right)
% \right]
%+\frac{\rho_{\rm P}}{\rho}\frac{\partial }{\partial {  r}_\beta}\left(\frac{m_{\rm B}}{\zeta\tau_{1,2}}{   \Theta}_{\alpha\beta}-n k_{\rm B} T\delta_{\alpha\beta} \right),\\
=\frac{\partial  }{\partial{ {   r}_\beta}}\left(\rho_{\rm P}({\pmb r},t)\frac{k_{\rm B}T}{\zeta}{  \delta}_{\alpha\beta}\right)
+\underbrace{ \frac{\rho_{\rm P}}{\rho}\frac{\partial }{\partial {  r}_\beta}\left(\frac{m_{\rm B}}{\zeta\tau_{1,2}}{   \Theta}_{\alpha\beta}-n k_{\rm B} T\delta_{\alpha\beta} \right)}_{\Xi}
\end{split}
\end{align}
In the dilute limit $\rho_{\rm P}/\rho\to 0$, therefore the term $\Xi\to 0$. In terms of number density ($\rho_{\rm P}=2 n_{\rm P} m_{\rm B}$), we get 
% \begin{align}
%  \begin{split}
% - \int d{\pmb Q } {  J}^{\rm r(1)}_{\alpha} 
% &=\frac{\partial  }{\partial{ {   r}_\beta}}\left(\rho_{\rm P}({\pmb r},t)\frac{k_{\rm B}T}{m_{\rm B}}{  \delta}_{\alpha\beta}\right).
% \end{split}
% \end{align}
%Finally,
\begin{align}
 \begin{split}
 \frac{\partial }{\partial t}  {n_{\rm P}}({\pmb r},  t)  + &  
\frac{\partial  }{\partial{\pmb   r }_\alpha } \left[
 n_{\rm P}\,U_\alpha+\frac{1}{2\zeta}\frac{\partial  }{\partial{    r }_\beta}{ \Theta}_{\alpha\beta} \right]
=\frac{k_{\rm B}T}{\zeta}\frac{\partial^2  n_{\rm P}}{\partial{ {  \pmb r}^2}}, 
\end{split}
\end{align}
which is the  required density equation \citep{beris1994compatibility,ottinger1996kinetic,apostolakis2002stress}.
 \bibliographystyle{unsrt}
 \bibliography{reference_new}
\end{document}